\documentclass[pra,twocolumn,showpacs,letterpaper,showpacs,superscriptaddress]{revtex4}
\usepackage{graphicx,amsmath,amssymb,amsfonts,latexsym,color,dcolumn,bm,subfigure}

\usepackage{amsmath}
\usepackage{amssymb}
\usepackage{graphicx}
\usepackage{stackrel}
\usepackage{color}

\renewcommand{\imath}[0]{\mathrm{i}}

\begin{document}

\title{A quasi-analytical modal approach for computing Casimir interactions in periodic nanostructures}

\author{F. Intravaia}
\affiliation{Theoretical Division, MS B213, Los Alamos National Laboratory, Los Alamos, NM 87545, USA}

\author{P.S. Davids}
\affiliation{Applied Photonics and Microsystems, Sandia National Laboratories, Albuquerque, NM 87185, USA}

\author{R.S. Decca}
\affiliation{Department of Physics, Indiana University-Purdue University Indianapolis, Indianapolis, Indiana 46202, USA}

\author{V.A. Aksyuk}
\affiliation{Center for Nanoscale Science and Technology, National Institute of Standards and Technology, Gaithersburg, Maryland 20899, USA}

\author{D. L\'{o}pez}
\affiliation{Center for Nanoscale Materials, Argonne National Laboratory, Argonne, Illinois 60439, USA}

\author{D.A.R. Dalvit}
\affiliation{Theoretical Division, MS B213, Los Alamos National Laboratory, Los Alamos, NM 87545, USA}

\begin{abstract}
We present an almost fully analytical technique for computing Casimir interactions between periodic lamellar gratings based on a modal approach.
Our method improves on previous work on Casimir modal approaches for nanostructures \cite{Davids10} by using the exact form of the eigenvectors
of such structures, and computing eigenvalues by solving numerically a simple transcendental equation. In some cases eigenvalues can be solved for
exactly, such as the zero frequency limit of gratings modeled by a Drude permittivity. Our technique also allows us to predict analytically the behavior of the Casimir interaction in limiting cases, such as the large separation asymptotics. The method can be generalized to more complex grating structures, and may provide
a deeper understanding of the geometry-composition-temperature interplay in Casimir forces between nanostructures. 
\end{abstract}

\maketitle

\section{Introduction}

Geometry, material composition, and temperature can strongly influence the Casimir interaction \cite{Casimir48} between objects separated by micron and sub-micron gaps. Recent theoretical developments have shown how to compute the Casimir force between complex structures using a variety of methods 
\cite{Buscher04,Lambrecht08,Johnson11,Rodriguez11}. Among these, we mention 
techniques based on the summation of zero-point energies \cite{Ninham70,Dalvit06,Intravaia07}, which are suitable for high symmetry problems; the scattering approach which requires the 
computation of the reflection matrices of the scatterers \cite{Lambrecht06,Rahi09,Rahi11,Lambrecht11a}; and full-wave numerical techniques, that compute the force from the Maxwell stress tensor \cite{Rodriguez09,Johnson11,Rodriguez11}.

In a previous paper \cite{Davids10} a modal approach was proposed to calculate finite-temperature Casimir interactions between 2D periodically modulated surfaces. This method uses the scattering formula for the Casimir free energy and computes the reflection amplitudes of the scatterers by decomposing the electromagnetic field into their natural modes. The modal approach is based on a plane-wave expansion of the fields and a Fourier
decomposition of the spatial-dependent permittivity of the structures, in the same way as done in rigorous coupled wave approaches (RCWA) in classical
photonics \cite{Busch07}. The modal method is limited to periodic structures, such as photonic crystals and metamaterials.  While other more general numerical scattering techniques exist, the modal expansion provides insight into the different (photonic, plasmonic, etc) mode contributions to the Casimir force,
thereby allowing to unveil otherwise hidden 
balances \cite{Intravaia05,Intravaia07,Haakh10}. Other RCWA techniques, not based on modal methods, have been
also used by the Casimir community to study Casimir forces \cite{Lambrecht08} and nanoscale heat transfer \cite{Guerout12} in grating structures.

In \cite{Davids10} both the eigenmodes and their eigenfrequencies were computed numerically  by solving  a non-self-adjoint eigenvalue problem \cite{Cole68,Naimark68}. In this paper we improve this previous work by developing an almost fully analytical modal approach to compute Casimir interactions between 1D lamellar grating structures, which is a generalization to Casimir physics of well-developed methods in grating theory \cite{Botten81b,Li93}. The key feature of our method is that the eigenmodes of the grating
can be solved for analytically without any Fourier expansion of the permittivity, while the eigenfrequencies are solutions to a simple transcendental
equation. Analytical expressions for the eigenfrequencies can be found in some limiting cases, such as for perfectly reflecting gratings, and for the low
frequency limit of real material gratings, described by simple Drude or plasma permittivities. The quasi-analytical modal approach also allows us to exactly demonstrate some properties of the scattering operators and to derive expressions for the Casimir interaction in some limiting cases, such as the large distance/low frequency limit, and the behavior of the force at high temperatures. The method can be generalized to more complex structures beyond 1D lamellar grating, and can also provide a detailed framework for the analysis of other fluctuation-induced interactions in nanostructures, including thermal emission and near-field heat transfer.

The general set up is similar to that of  \cite{Davids10}, which we briefly outline here.  Within the framework of the scattering approach, the calculation of the Casimir free energy 
\begin{equation}
\mathcal{F}(a)= \frac{1}{\beta} \sum^{\infty'} _{l=0}  {\rm Tr} \log  \left[1- \underleftarrow{\mathcal{R}}^{L} \cdot \underrightarrow{\mathcal{X}}(a)
\cdot \underrightarrow{\mathcal{R}}^{R} \cdot \underleftarrow{\mathcal{X}}(a) \right] ,
\label{freeEnergy}
\end{equation}
is essentially reduced to the calculation of the scattering matrices of isolated objects.
The symbol ${\rm Tr}$ indicates the trace over spacial and polarization degrees of freedom \cite{Davids10}.
Here $\beta=1/k_{\rm B} T$ is the inverse temperature, $\mathcal{X}$ represents translation matrices that depend of the distance $a$ between the gratings, and $\cal{R}$ are their reflection matrices. All these matrices are evaluated at the Matsubara imaginary frequencies $\omega_l=i \xi_{l}=i 2\pi l k_{B}T/\hbar$ \cite{Matsubara55}, and the primed sum indicates that the $l=0$ term has half weight. The arrows under the reflection and translation matrices indicate the direction of propagation of light - for example, $\underleftarrow{\mathcal{R}}^{L}$ is the reflection on the left grating for light propagating from left to right. The translation matrices are diagonal in a plane-wave, Rayleigh basis (see \cite{Davids10} for explicit expressions). 

Our goal in the rest of the paper is to compute the reflection matrix of an isolated 1D lamellar grating with the quasi-analytical modal technique.
\begin{figure}
\centering\includegraphics[width=6.5cm]{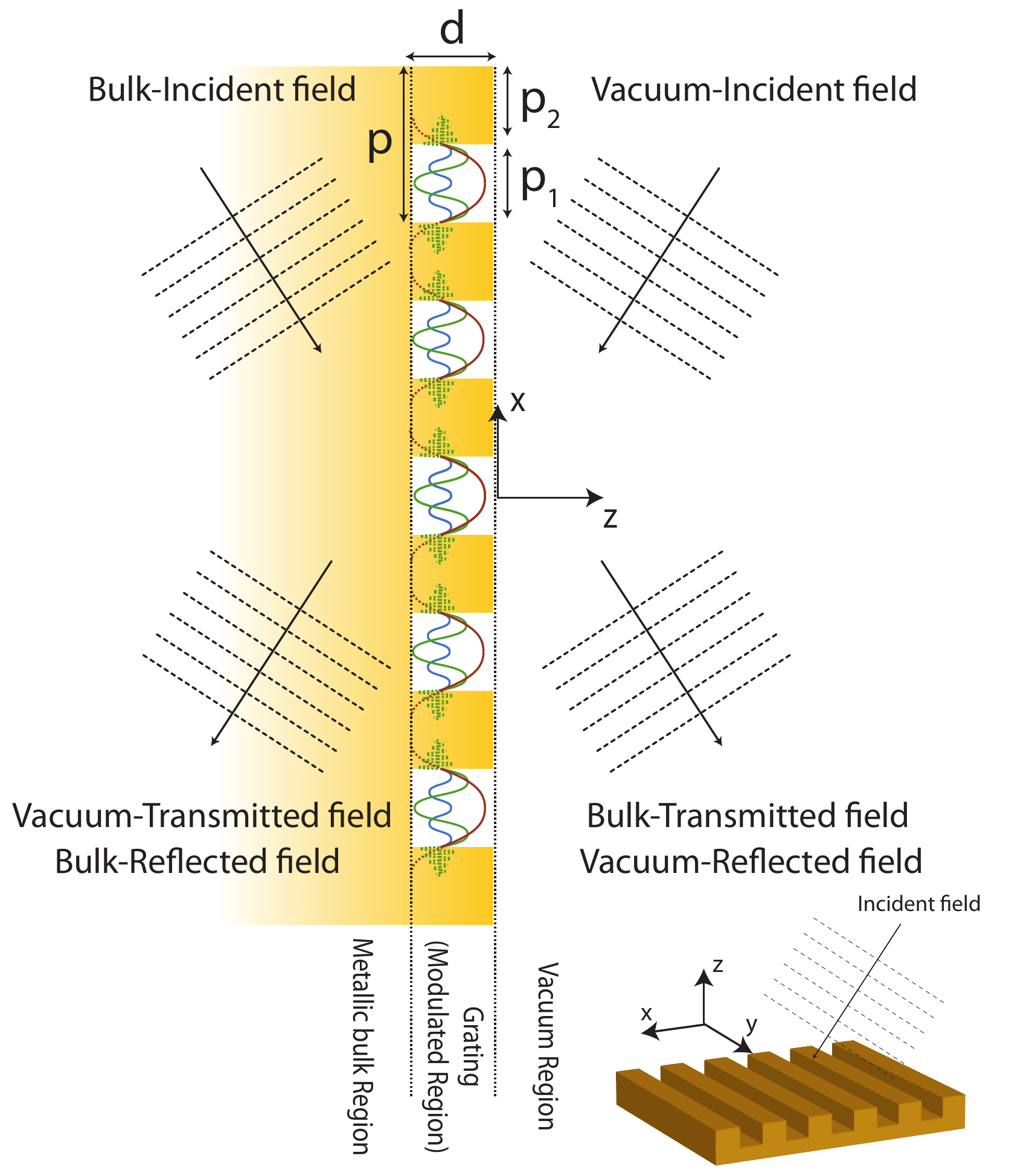}
\caption{ A schematic representation of the scattering of the electromagnetic field on a 1D grating.}
\label{fig:gratings}
\end{figure}
In the following we will analyze the scattering properties of a 1D lamellar grating of depth $d$ and period $p=p_{1}+p_{2}$, where $p_{1}$ is the width of the grooves and $p_{2}$ the with of the teeth (see Fig.\ref{fig:gratings}).
We divide the 1D lamellar grating into three regions: (i) the homogeneous, vacuum region above the grating, $z>0$; (ii) the region $z<-d$ below the grating, filled with homogeneous medium of permittivity $\epsilon(\omega)$ and permeability $\mu(\omega)$; and (iii) the grating region $-d<z<0$, where the space is filled with the modulated medium $\epsilon(x;\omega)$ and $\mu(x;\omega)$ describing the 1D lamellar grating. In each $i$-th region 
($i=v$, vacuum region; $i=g$, grating region; and $i=m$, bulk medium region),
the solution of Maxwell equation can be written as
\begin{eqnarray}
\mathbf{F}^{(i)}(x,z)
&=&
\begin{pmatrix}
E_{x}(x,z)\\
E_{y}(x,z)\\
H_{x}(x,z)\\
H_{y}(x,z)
\end{pmatrix}^{(i)} \nonumber \\
&=&\sum_{\nu,s}A_{\nu}^{(s,i)}\mathbf{Y}^{(s,i)}[x,\lambda^{(s,i)}_{\nu}] e^{\imath\lambda_{\nu}^{(s,i)}z} ,
\label{FieldDecomposition}
\end{eqnarray}
where we have considered the four independent transverse field components (the remaining two components can be directly calculated from the previous four). Moreover, since the system is invariant with respect to translations along the $y$-direction, eigenmodes have a $e^{i k_{y} y}$ plane wave
form, which we omitted in the above equation. $\mathbf{Y}^{(s,i)}[x,\lambda^{(s,i)}_{\nu}]$ is a column vector describing the $x$ dependence of the eigenvector 
with corresponding eigenvalue $\lambda^{(s,i)}_{\nu}$, where $\nu$ labels the eigenvalue and $s$ denotes one of the two possible polarizations. 
$A_{\nu}^{(s,i)}$ are complex amplitudes, to be determined by imposing the following boundary conditions at the interfaces (continuity of the tangental components of ${\bf E}$ and ${\bf H}$ at the interfaces):
\begin{widetext}
\begin{subequations}
\begin{gather}
\sum_{\nu,s}A_{\nu}^{(s,g)}\mathbf{Y}^{(s,g)}[x,\lambda_{\nu}]=\sum_{\nu,s}A_{\nu}^{(s,v)}\mathbf{Y}^{(s,v)}[x,\lambda_{\nu}] ,  \\
\sum_{\nu,s}A_{\nu}^{(s,m)}\mathbf{Y}^{(s,m)}[x,\lambda_{\nu}] e^{-\imath\lambda_{\nu}^{(s,m)}d}=\sum_{\nu,s}A_{\nu}^{(s,g)}\mathbf{Y}^{(s,g)}[x,\lambda_{\nu}] e^{-\imath\lambda_{\nu}^{(s,g)}d} ,
\end{gather} 
\label{boundary-equations}
\end{subequations}
\end{widetext}
where we have dropped the superscript for the eigenvalues argument of the eigenvectors because they are the same as the eigenvectors. 
Using properties of the eigenvectors (see below) it is possible to derive the $A_{\nu}^{(s,v)}$ in terms of the $A_{\nu}^{(s,m)}$ or vice-versa,
and then extract the scattering operators of the grating.

The expressions for the eigenvectors and the corresponding eigenvalues play a central role in our derivation. In the following we will focus on the evaluation of these quantities for the grating region ($i=g$).  The results will be also valid for the bulk and vacuum homogeneous regions. For this one simply needs
to take the limit of no modulation ($\epsilon(x;\omega)=\epsilon(\omega)$ and 
$\mu(x;\omega)=\mu(\omega)$).


\section{Non-self-adjoint eigenvalue problem}

In this section we  give the mathematical details on how to solve Maxwell equations in a 1D modulated magneto-dielectric region. 
The modulation is lamellar along the $x-$ direction (the electric permittivity and the magnetic permeability modulated along that direction), and the grating is invariant along the $y-$ direction. For the purposes of finding the eigenmodes in the grating region, we assume that the system is also invariant
along the $z-$ direction \cite{Botten81b,Li93}. 
Using the invariance in $y$, it is possible to write Maxwell equations 
$\nabla \times \mathbf{E}-\imath \omega \mu\mathbf{H}=0$ and
$\nabla \times \mathbf{H}+\imath \omega \epsilon\mathbf{E}=0$ in the following form
\begin{subequations}
\begin{align}
&\partial_{z} E_{x}= -\partial_{x}\left(\frac{k_{y}}{\omega\epsilon}H_{x}\right) +
\left[\partial_{x}\frac{1}{\imath\omega \epsilon }\partial_{x}-\imath \omega\mu\right]H_{y} , \\
&\partial_{z} E_{y}=-\frac{\tilde{k}^{2}}{\imath \omega\epsilon}H_{x} +  \frac{k_{y}}{\omega\epsilon}\partial_{x}H_{y} , \\
&\partial_{z} H_{x}= \partial_{x}\left(\frac{k_{y}}{\omega\mu}E_{x}\right)  -\left[\partial_{x}\frac{1}{\imath\omega \mu}\partial_{x}-\imath \omega\mu\right]E_{y} , \\
&\partial_{z} H_{y}= \frac{\tilde{k}^{2}}{\imath\omega\mu}E_{x}  -\frac{k_{y}}{\omega\mu}\partial_{x}E_{y},
\end{align}
\end{subequations}
where we already eliminated the $z$-component of the electric and the magnetic fields. 
For the sake of simplicity we will also be measuring all frequencies as wave vectors,  so that $\omega/c \to \omega$.

The previous system of equations can be solved by separation of variables, by writing
\begin{equation}
\begin{pmatrix}
E_{x}(x,z)\\
E_{y}(x,z)\\
H_{x}(x,z)\\
H_{y}(x,z)
\end{pmatrix} 
=
\begin{pmatrix}
E_{x}[x,\lambda]\\
E_{y}[x,\lambda]\\
H_{x}[x,\lambda]\\
H_{y}[x,\lambda]
\end{pmatrix} 
 e^{i \lambda z}
\equiv
{\bf Y}[x,\lambda] e^{i \lambda z} ,
\end{equation}
where ${\bf Y}[x,\lambda]$ satisfies

\begin{widetext}

\begin{equation}
\lambda
{\bf Y}[x,\lambda]=
\begin{pmatrix}
0&0 &-\frac{\partial_{x}}{\imath}\frac{k_{y}}{\omega\epsilon} &-\left[\partial_{x}\frac{1}{\omega\epsilon}\partial_{x}+\omega\mu\right] 
\vspace{2mm} 
\\
0&0& \frac{\tilde{k}^{2}}{\omega\epsilon} &\frac{k_{y}}{\omega\epsilon}\frac{\partial_{x}}{\imath}\vspace{2mm}  \\
\frac{\partial_{x}}{\imath}\frac{k_{y}}{\omega\mu}& \left[\partial_{x}\frac{1}{\omega\mu}\partial_{x}+\omega\epsilon\right]&0&0 \vspace{2mm} \\
-\frac{\tilde{k}^{2}}{\omega\mu}& -\frac{k_{y}}{\omega\mu}\frac{\partial_{x}}{\imath}&0&0
\end{pmatrix}
{\bf Y}[x,\lambda] ,
\label{vectEquation}
\end{equation}
\end{widetext}
which is a eigenvalue equation for the eigenvector ${\bf Y}$ with eigenvalues $\lambda$.
Here $\tilde{k}^{2}=k^{2}-k_{z}^{2}$ and $k^{2}=\mu\epsilon \omega^{2}$, and for simplicity we have omitted the spatial and frequency dependency of the permittivity and permeability functions. 

The $4\times 4$ matrix equation (\ref{vectEquation}) can be easily transformed into a $2\times 2$ second order differential equation either in the $E$- or $H$-components only. As a further simplification we decompose the fields in two independent polarizations. We will define ``$e$'' or ``$h$''  polarizations, for which the $x$-component of the electric or magnetic field vanishes respectively, i.e. $E^{e}_{x}=0$ and $H^{h}_{x}=0$.
Using this decomposition it is possible to show  that the $2\times 2$ matrix equation decouples into two one dimensional second order (in general non-self-adjoint) differential equations for the $y$-components of the fields, namely
\begin{equation}
\left[\sigma^{(s)}(x)\partial_{x}\frac{1}{\sigma^{(s)}(x)} \partial_{x}+\tilde{k}^{2}(x)\right]\mathcal{U}^{(s)}[x,\lambda]=\lambda^{2}\mathcal{U}^{(s)}[x,\lambda] ,
\label{Li1}
\end{equation}
where $s=e,h$, $\sigma^{(e)}(x)=\mu(x)$, $\sigma^{(h)}(x)=\epsilon(x)$, 
$\mathcal{U}^{(e)}=E_y^{(e)}$, and $\mathcal{U}^{(h)}=H_y^{(h)}$.
The corresponding eigenvalue $\lambda$ will therefore also depend on the polarization $s$. 

Given Eq.(\ref{Li1}), from Maxwell equations we get  \cite{Li93}
\begin{equation}
J^{(s)}[x,\lambda]=\frac{\delta^{(s)}}{i \lambda \sigma^{(s)}(x)} \frac{k_{y}}{\omega}\partial_{x} \mathcal{U}^{(s)}[x,\lambda] ,
\label{Li3}
\end{equation}
where $\delta^{(e)}=-1$,  $\delta^{(h)}=1$, $J^{(e)}=H_y^{(e)}$, and $J^{(h)}=E_x^{(h)}$. 
Solving (\ref{Li1}) and using the solutions in (\ref{Li3}) one can therefore find the $y$ components of the eigenvector ${\bf Y}$, and from them, using again 
Maxwell equations, the $x$ components, given by
\begin{equation}
\begin{pmatrix}
E_x\\
H_x
\end{pmatrix}
=\frac{\imath}{\tilde{k}^2}
\begin{pmatrix}
k_{y}\partial_{x}& i \lambda \mu \omega \\
- i \lambda \epsilon \omega &k_{y}\partial_{x}
\end{pmatrix}
\begin{pmatrix}
E_y\\
H_y
\end{pmatrix} .
\end{equation}
Finally one can write the eigenvectors for the two polarizations
\begin{gather}
\mathbf{Y}^{(e)}[x,\lambda]
=
\begin{pmatrix}
0\\
\mathcal{U}^{(e)}[x,\lambda]\\
\frac{\lambda^{2}+k_{y}^{2}}{  \omega\lambda}\frac{\mathcal{U}^{(e)}[x,\lambda]}{\mu(x)} \\
-\frac{k_{y}}{\omega\lambda}\frac{ \partial_x \mathcal{U}^{(e)}[x,\lambda]}{\imath\mu(x)}
\end{pmatrix},
\nonumber\\
\mathbf{Y} ^{(h)}[x,\lambda]=
\begin{pmatrix}
-\frac{\lambda^{2}+k_{y}^{2}}{\omega\lambda}\frac{\mathcal{U}^{(h)}[x,\lambda]}{\epsilon(x)}\\
\frac{k_{y}}{ \omega\lambda}\frac{\partial_x \mathcal{U}^{(h)}[x,\lambda]}{\imath\epsilon(x)}\\
0 \\
\mathcal{U}^{(h)}[x,\lambda]
\end{pmatrix} .
\label{EigenVectors}
\end{gather}

The matrix differential operator in the r.h.s of Eq.(\ref{vectEquation})
is clearly not Hermitian. Therefore the eigenvalue problem \eqref{vectEquation} is non-self-adjoint and the eigenvalues $\lambda$ are in general complex \cite{Cole68,Naimark68}. Note that, since the matrix is non-symmetric, this remains true even if we consider a non dissipative material. The existence of such complex values is associated with the presence of evanescent fields in the structure \cite{Carniglia71}.

Following the theory of non-self-adjoint differential equations \cite{Cole68,Naimark68},
one also needs to find the adjoint eigenvectors, which are bi-orthogonal to the eigenvectors in Eq.(\ref{EigenVectors}), in order to completely characterize the mathematical description.
Indicating them with $\overline{\bf Y}[x,\lambda]$, one can show that they are solutions to

\begin{widetext}
\begin{equation}
\lambda
\overline{\bf Y}[x,\lambda]=
\begin{pmatrix}
0&0 & \frac{k_{y}}{\omega\mu^*} \frac{\partial_{x}}{\imath}  & -\frac{\tilde{k}^{*2}}{\omega\mu^*}
\vspace{2mm} 
\\
0&0& \left[\partial_{x}\frac{1}{\omega\mu^*}\partial_{x}+\omega\epsilon^*\right] &- \frac{\partial_{x}}{\imath} \frac{k_{y}}{\omega\mu^*} \vspace{2mm}  \\
- \frac{k_{y}}{\omega\epsilon^*} \frac{\partial_{x}}{\imath}   &  \frac{\tilde{k}^{*2}}{\omega\epsilon^*}  &0&0 \vspace{2mm} \\
 -\left[\partial_{x}\frac{1}{\omega\epsilon^*}\partial_{x}+\omega\mu^*\right]  & \frac{\partial_{x}}{\imath}  \frac{k_{y}}{\omega\epsilon^*} &0&0
\end{pmatrix}
\overline{\bf Y}[x,\lambda] .
\label{vectAdjointEquation}
\end{equation}
\end{widetext}
The two eigenvalue problems \eqref{vectEquation} and \eqref{vectAdjointEquation} have the same eigenvalue spectrum \cite{Cole68,Naimark68}.
One can find the adjoint eigenvectors employing the same strategy used above. One gets
\begin{gather}
\overline{\mathbf{Y}}^{(e)}[x,\lambda]
=
\begin{pmatrix}
\frac{k_y}{\lambda^2+k_y^2} \frac{ \partial_x \mathcal{V}^{(e)}[x,\lambda]}{\imath\mu^*(x)}\\
\frac{\mathcal{V}^{(e)}[x,\lambda]}{\mu^*(x)} \\
\frac{\omega \lambda}{\lambda^2+k_y^2}  \mathcal{V}^{(e)}[x,\lambda] \\
0
\end{pmatrix},
\nonumber\\
\overline{\mathbf{Y}} ^{(h)}[x,\lambda]=
\begin{pmatrix}
- \frac{\omega \lambda}{\lambda^2+k_y^2} \mathcal{V}^{(h)}[x,\lambda] \\
0 \\
\frac{k_y}{\lambda^2+k_y^2} \frac{\partial_x \mathcal{V}^{(h)}[x,\lambda]}{\imath\epsilon^*(x)}\ \\
\frac{\mathcal{V}^{(h)}[x,\lambda]}{\epsilon^*(x)} 
\end{pmatrix} ,
\label{EigenAdjointVectors}
\end{gather}
where the function $\mathcal{V}^{(s)}[x,\lambda]$ satisfies the differential equation
\begin{equation}
\left[[\sigma^{(s)}(x)]^{*}\partial_{x}\frac{1}{[\sigma^{(s)}(x)]^{*}}\partial_{x}+\tilde{k}^{*2}\right] \mathcal{V}^{(s)}[x,\lambda]=
\lambda^2 \mathcal{V}^{(s)}[x,\lambda] ,
\end{equation}
which is the adjoint equation of \eqref{Li1}. The eigenvectors and their adjoint are bi-orthogonal, i.e.
\begin{eqnarray}
\langle 
\overline{\mathbf{Y}}^{(s)}[\lambda_{\gamma}] |
\mathbf{Y}^{(s')}[\lambda_{\gamma'}]
\rangle
&\equiv&
\int_{-p/2}^{p/2} dx \frac{ {\overline{\bf Y}}^{(s)}[x,\lambda_{\gamma}]^{\dagger} \cdot {\bf Y}^{(s')}[x,\lambda_{\gamma'}]} {\sigma^{(s)}(x) }
\nonumber \\
&=&
\delta_{\gamma,\gamma'} \delta_{s,s'} .
\label{biorthogonality2}
\end{eqnarray}
In deriving the previous expression we used that the functions $\mathcal{U}$ and $\mathcal{V}$ satisfy themselves the bi-orthogonality 
relation 
\begin{equation}
\int_{-p/2}^{p/2} dx \frac{\mathcal{V}^{(s)*}[x,\lambda_{\gamma}] \mathcal{U}^{(s')}[x,\lambda_{\gamma'}] }{\sigma^{(s)}(x)} = 
\frac{1}{2} \delta_{\gamma,\gamma'}  \delta_{s.s'} .
\end{equation}
The choice of the normalization factor of $\mathcal{V}^{(s)}$ and $\mathcal{U}^{(s)}$ allows us to have
$\mathcal{V}^{(s)*}[x,\lambda]= \mathcal{U}^{(s)}[-x,\lambda]$, which will be particularly handy for the forthcoming evaluations \cite{Li93}. 

It is interesting to note that, from the symmetry in the equations in \eqref{vectEquation} and \eqref{vectAdjointEquation}, one can also show that the bi-orthogonality relation \eqref{biorthogonality2} also has a physical meaning. It is  directly connected with 
Poynting vector reciprocity theorem, and therefore with the energy flux along the $z$ direction.


\section{Eigenmodes and eigenvalues}

The approach described in the previous section requires the solution of the second order differential equation given in Eq. \eqref{Li1}. Because of the periodicity of
our system the functions $\mathcal{U}^{(s)}$ and their derivatives must satisfy pseudo-periodic (Bloch-Floquet) boundary conditions, i.e.
$ \mathcal{U}^{(s)}[p/2,\lambda]=e^{\imath \alpha_{0}p}\mathcal{U}^{(s)}[-p/2,\lambda]$ and
$\partial_x \mathcal{U}^{(s)}[p/2,\lambda]=e^{\imath \alpha_{0}p} \partial_x \mathcal{U}^{(s)}[-p/2,\lambda]$.  
The previous requirements define a (in general non-self-adjoint) scalar eigenvalue problem.
In this section we give the form of the eigenfunctions $\mathcal{U}^{(s)}[x,\lambda]$ and the 
corresponding eigenvalues $\lambda$, both for the grating and homogeneous regions. 

\subsection{Eigenfunctions in grating region}

Within the modulated region ($-d<z<0$) of the 1D lamellar grating, the permittivity and permeability functions are 
\begin{subequations}
\begin{gather}
\epsilon(x)=\begin{cases}
\epsilon_{1}& {\rm for}\;  |x|\le \frac{p_{1}}{2}\\
\epsilon_{2}& {\rm for} \; \frac{p_{1}}{2}\le |x|\le \frac{p}{2}
\end{cases}
\\
\mu(x)=
\begin{cases}
\mu_{1}& {\rm for}\;  |x|\le \frac{p_{1}}{2}\\
\mu_{2}&  {\rm for} \; \frac{p_{1}}{2}\le |x|\le \frac{p}{2} .
\end{cases}
\end{gather}
\end{subequations}
Using these expressions in the differential equation \eqref{Li1} one can find its solutions $\mathcal{U}^{(s)}[x,\lambda]$ must have the following expression
\begin{widetext}
\begin{subequations}
\begin{equation}
\mathcal{U}^{(s)}[x,\lambda]=
\frac{C^{(s)}(\lambda)}{2}\left\{ \cos( \alpha_{0} p/2) \varphi^{(s)}_{e}[x,\lambda]+\imath \sin(\alpha_{0} p/2) \varphi^{(s)}_{o}[x,\lambda]\right\},
\label{definition-U}
\end{equation}
where $\alpha_{0}$ describes the $x$ component of the wave vector limited within the first Brilloin zone ($-\pi/p\le k_{x}=\alpha_{0}\le \pi/p$),
$\varphi^{(s)}_{e}[x,\lambda]=\phi^{(s)}_{e}[x,\lambda]/\phi^{(s)}_{e}[p/2,\lambda]$ and
$\varphi^{(s)}_{o}[x,\lambda]=\phi^{(s)}_{o}[x,\lambda]/\phi^{(s)}_{o}[p/2,\lambda]$.
The functions $\phi^{(s)}_{e}[x,\lambda]$ and $\phi^{(s)}_{o}[x,\lambda]$, even and odd in the variable $x$ respectively, are given by
\begin{eqnarray}
\phi^{(s)}_{e}[x,\lambda]
&=&
\begin{cases}
\cos(\gamma_{1}x) , \quad {\rm for} \quad |x|\le \frac{p_{1}}{2}\\
\cos\left(\gamma_{1}\frac{p_{1}}{2}\right) \cos\left(\gamma_{2} \left[|x|-\frac{p_{1}}{2}\right]\right) - \frac{\sigma^{(s)}_{2}\gamma_{1}}{\sigma^{(s)}_{1}\gamma_{2}}
\sin\left(\gamma_{1}\frac{p_{1}}{2}\right) \sin\left(\gamma_{2}\left[|x|-\frac{p_{1}}{2}\right] \right)  , \quad {\rm for} \quad \frac{p_{1}}{2}\le |x|\le 
\frac{p}{2}
\end{cases}
\label{function-even}
\\
\phi^{(s)}_{o}[x,\lambda] 
&=&
\begin{cases}
\frac{\sin\left(\gamma_{1}x\right)}{\gamma_{1}} , \quad {\rm for} \quad |x|\le \frac{p_{1}}{2}\\
\mathrm{sign}(x) \frac{\sin\left(\gamma_{1}\frac{p_{1}}{2}\right) \cos\left( \gamma_{2}\left[|x|-\frac{p_{1}}{2}\right] \right) +
\frac{\sigma^{(s)}_{2}\gamma_{1}}{\sigma^{(s)}_{1}\gamma_{2}}
\cos\left(\gamma_{1}\frac{p_{1}}{2}\right) \sin\left[\gamma_{2}\left(|x|-\frac{p_{1}}{2}\right)\right]}{\gamma_{1}} , \quad {\rm for} \quad  \frac{p_{1}}{2}\le |x|\le 
\frac{p}{2}
\end{cases}
\label{function-odd}
\end{eqnarray}
where $\gamma^{2}_{1}=\epsilon_1 \mu_1 \omega^2-(k^{2}_{z}+\lambda^{2})$ and $\gamma^{2}_{2}=\epsilon_2 \mu_2  \omega^2-(k^{2}_{z}+\lambda^{2})$.
The normalization constant
$C^{(s)}(\lambda)$ is given by
\begin{equation}
C^{(s)}(\lambda)=\left[\cos^2(\alpha_{0}p /2) \int^{\frac{p}{2}}_{0}{\rm d}x \frac{\varphi^{(s)}_{e}[x,\lambda]^{2}}{\sigma^{(s)}(x)}+
\sin^2(\alpha_{0} p /2) \int^{\frac{p}{2}}_{0}{\rm d}x \frac{\varphi^{(s)}_o[x,\lambda]^{2}}{\sigma^{(s)}(x)}\right]^{-\frac{1}{2}} .
\label{normalization}
\end{equation}
\label{eigenfunctions}
\end{subequations}
\end{widetext}
We note that all the above functions, being even in $\gamma_{1,2}$,  do not depend on the definition (sign) of the square root, i.e.,  do not contain any branch cut.
We also note that  $\mathcal{U}^{(s)}[x,\lambda]=\mathcal{U}^{(s)}[x,-\lambda]$.  
The previous expressions \eqref{eigenfunctions} are fully determined only once the eigenvalues $\lambda$ are known. 


\subsection{Eigenvalues in the grating region- General properties}

Imposing the pseudo-periodic boundary conditions (on the function and its derivative) it is possible to show that the eigenvalues are the solution of the following transcendental equation \cite{Li93}

\begin{eqnarray}
0= D^{(s)}(\lambda) &\equiv & -\cos(\alpha_{0} p) + \cos(p_{1} \gamma_1)\cos(p_{2}\gamma_{2}) \nonumber \\
&& -T^{(s)}(\lambda) \sin(p_{1}\gamma_{1}) \sin(p_{2}\gamma_{2})  ,
\label{TrEq}
\end{eqnarray}
where
\begin{gather}
T^{(s)}(\lambda)=\frac{1}{2}
\left(\frac{\sigma^{(s)}_{1}\gamma_{2}}{\sigma^{(s)}_{2}\gamma_{1}}+\frac{\sigma^{(s)}_{2}\gamma_{1}}{\sigma^{(s)}_{1}\gamma_{2}}\right).
\end{gather}
Eq \eqref{TrEq} clearly shows that the eigenvalues depend on the frequency $\omega$, and the two wave-vectors $\alpha_0$ and $k_y$. 
We can deduce the following properties, valid for both polarizations (we will drop the superscript $s$):
\begin{itemize}
\item  $D(\lambda)$ is quadratic in $\lambda$ and, therefore, if $\lambda$ is a solution then $-\lambda$ is also a solution. 
\item  $D(\lambda)$ is even in $\gamma_{i}$, which implies that the solution is not affected by the sign of the square root.
\item  $D(\lambda)$ is even in $\alpha_{0}$ and $k_{y}$, which implies that  $\lambda$ is an even function of the two variables.
\item For complex frequencies $\omega=\zeta$,  $\lambda(\zeta)=-\lambda^*(-\zeta^{*})$, which implies that $\lambda$ is a pure imaginary quantity for imaginary frequencies $\omega=i \xi$.
\item At high frequencies the ultraviolet  transparency of all materials ($\epsilon=\mu=1$ for $\omega \to \infty$) implies that
\begin{equation}
 \lambda(\omega \rightarrow \infty)  = \pm\sqrt{\omega^2- \left[k_{y}^{2}+\left(\alpha_{0}+\frac{2\pi \nu}{p}\right)^{2} \right]},
\end{equation}
with $\nu \in \mathbb{Z}$ (see also Section \ref{DPeigen}). 
\item The solutions of $D(\lambda)=0$ form an infinite, numerable set of complex numbers.
\end{itemize}


\subsection{Special case: homogeneous media}

The previous formalism also applies  to the homogeneous region of space. This particular limit is reached by imposing $\sigma_{1}=\sigma_{2}\equiv\sigma$ in the previous expressions, which implies $\gamma_{1}=\gamma_{2}\equiv\gamma$. 
The corresponding transcendental equation becomes then
\begin{equation}
D(\lambda)=
\cos(\gamma p)-\cos(\alpha_{0} p) =0 .
\end{equation}
There are  two possible sets of solutions for $\gamma$, namely $\gamma=  \pm \alpha_{0}+ \frac{2\pi \nu}{p} \equiv
\alpha_{\nu,\pm}$, where $\nu \in \mathbb{Z}$. The corresponding eigenvalues are
\begin{equation}
\lambda_{\nu}=\pm\sqrt{\epsilon\mu \omega^2-(k^{2}_{y}+\alpha_{\nu,\pm}^{2})} ,
\label{eigenvalues-homogeneous}
\end{equation}
and are the same for both polarizations $s=e$ and $s=h$.
From a comparison with the usual grating theory each value of $\nu$ corresponds to a specific Rayleigh order. However, it is important to note that 
for the eigenvalues $\lambda$ the set of solutions with the $+$ sign gives an identical result as the set of solutions with the $-$ sign. This means that, for $\nu=0,\pm 1,\pm 2,\dots$, one needs to consider either one or the other.
Instead, both set of solutions must be considered if we limit $\nu=0, 1, 2,\dots$ (the $\nu=0$ solution must be counted only once).
The eigenfunctions are 
\begin{equation}
\mathcal{U}^{(s)}[x,\lambda_{\nu}]=(-1)^{\nu}\sqrt{\frac{\sigma^{(s)}}{2p}}e^{\imath\left(|\alpha_{0}|+ \frac{2\pi \nu}{p}\right) x} ,
\label{eigenfunctions-homogeneous}
\end{equation}
which are the usual plane wave Rayleigh modes.  


\section{Scattering operators}

In this section we describe how to compute the reflection matrices of the nanostructure within the modal approach. What follows is
similar to what is presented in \cite{Davids10} with the important difference that now we have analytical expressions for the eigenvectors.
We introduce a transfer matrix that relates the field amplitudes of the vacuum and bulk regions, and express
the scattering operators of the grating in terms of the transfer matrix. 

\subsection{Transfer matrix of the grating}

We start by splitting the complex eigenvalues $\lambda$ into two subsets: (a) eigenvalues with positive real part, with corresponding eigenvectors
called ``right" eigenvectors, and (b) eigenvalues with negative real part, with corresponding ``left" eigenvectors. Each subset is then ordered
by the increasing moduli of the eigenvalues, the smallest eigenvalue denoted as $\lambda_{\nu=0}$ for the subset (a) [$-\lambda_{\nu=0}$
for subset (b)], $\lambda_{\nu=1}$ [$-\lambda_{\nu=1}$] for the next eigenvalues, etc. Although this ordering is not unique, the end results are not
affected by our choice. 

We define the fundamental $4\times \infty$ matrices
\begin{widetext}
\begin{eqnarray}
\underrightarrow{\mathcal{Y}}^{(i)} &=&
\begin{pmatrix}
|\mathbf{Y}^{(e)}[\lambda^{(e,i)}_{\nu=0}]\rangle, &|\mathbf{Y}^{(h)}[\lambda^{(h,i)}_{\nu=0}]\rangle, &
|\mathbf{Y}^{(e)}[\lambda^{(e,i)}_{\nu=1}]\rangle, &|\mathbf{Y}^{(h)}[\lambda^{(h,i)}_{\nu=1}]\rangle, &
\cdots 
\end{pmatrix},\nonumber\\
\underleftarrow{\mathcal{Y}}^{(i)} &=&
\begin{pmatrix}
|\mathbf{Y}^{(e)}[-\lambda^{(e,i)}_{\nu=0}]\rangle, &|\mathbf{Y}^{(h)}[-\lambda^{(h,i)}_{\nu=0}]\rangle, &
|\mathbf{Y}^{(e)}[-\lambda^{(e,i)}_{\nu=1}]\rangle, &|\mathbf{Y}^{(h)}[-\lambda^{(h,i)}_{\nu=1}]\rangle, &
\cdots
\end{pmatrix} ,
\end{eqnarray}
\end{widetext}
where we dropped the $x$ dependency of ${\bf Y}^{(s)}[x,\lambda]$. We recall that
the index $i$ indicates the region under consideration ($i=v,g,m$).
We also define the $\infty \times 4$ matrices formed by the adjoint eigenvectors
\begin{equation}
\underrightarrow{\mathcal{Y}}^{(i) \dag}=
\begin{pmatrix}
\langle \overline{\mathbf{Y}}^{(e)}[\lambda^{(e,i)}_{\nu=0}]|\\
\langle\overline{\mathbf{Y}}^{(h)}[\lambda^{(h,i)}_{\nu=0}]|\\
\langle\overline{\mathbf{Y}}^{(e)}[\lambda^{(e,i)}_{\nu=1}]|\\
\langle\overline{\mathbf{Y}}^{(h)}[\lambda^{(h,i)}_{\nu=1}]|\\
\vdots
\end{pmatrix},\quad
\underleftarrow{\mathcal{Y}}^{(i) \dag}=
\begin{pmatrix}
\langle\overline{\mathbf{Y}}^{(e)}[-\lambda^{(e,i)}_{\nu=0}]|\\
\langle\overline{\mathbf{Y}}^{(h)}[-\lambda^{(h,i)}_{\nu=0}]|\\
\langle\overline{\mathbf{Y}}^{(e)}[-\lambda^{(e,i)}_{\nu=1}]|\\
\langle\overline{\mathbf{Y}}^{(h)}[-\lambda^{(h,i)}_{\nu=1}]|\\
\vdots
\end{pmatrix} .
\end{equation}
Using the scalar product defined in \eqref{biorthogonality2} we have
$\underleftarrow{\underrightarrow{\mathcal{Y}}}^{(i) \dag} \cdot  \underleftarrow{\underrightarrow{\mathcal{Y}}}^{(i)}=\mathbb{I}$ and 
$\underrightarrow{\underleftarrow{\mathcal{Y}}}^{(i) \dag} \cdot  \underleftarrow{\underrightarrow{\mathcal{Y}}}^{(i)}=0$.
Let us define also the diagonal propagation matrices
\begin{align}
&\underrightarrow{\mathcal{P}}^{(i)}(z)=\mathrm{diag}[e^{\imath\lambda^{(e,i)}_{\nu=0}z},e^{\imath\lambda^{(h,i)}_{\nu=0}z},\cdots] ,
\nonumber\\
&\underleftarrow{\mathcal{P}}^{(i)}(z)=\mathrm{diag}[e^{-\imath\lambda^{(e,i)}_{\nu=0}z},e^{-\imath\lambda^{(h,i)}_{\nu=0}z},\cdots] ,
\label{ExpProp}
\end{align}
which clearly verify
$\underleftarrow{\underrightarrow{[\mathcal{P}}}^{(i)}(z)]^{-1}=\underrightarrow{\underleftarrow{\mathcal{P}}}(z)=\underleftarrow{\underrightarrow{\mathcal{P}}}(-z)$.
Using the previous definitions, the field in eq.\eqref{FieldDecomposition} can be written as
\begin{equation}
\mathbf{F}^{(i)}(x,z)=
\left(\underleftarrow{\mathcal{Y}}^{(i)},\underrightarrow{\mathcal{Y}}^{(i)}\right)
\cdot
\begin{pmatrix}
\underleftarrow{\mathcal{P}^{(i)}}&0\\
0&\underrightarrow{\mathcal{P}^{(i)}}
\end{pmatrix}
\cdot
\begin{pmatrix}
\underleftarrow{\mathbf{A}}^{(i)}\\
\underrightarrow{\mathbf{A}}^{(i)}
\end{pmatrix} ,
\end{equation}
where $\underleftarrow{\underrightarrow{\mathbf{A}}}^{(i)}$ is a column vector formed by the amplitudes
$A_{\nu}^{(s,i)}$ in \eqref{FieldDecomposition}.

By applying the boundary conditions (\ref{boundary-equations}), one gets the following relation between the amplitude coefficients 
\begin{equation}
\begin{pmatrix}
\underleftarrow{\tilde{\mathbf{A}}}^{(m)} \\
\underrightarrow{\tilde{\mathbf{A}}}^{(m)}
\end{pmatrix}=
\Theta
\cdot
\begin{pmatrix}
\underleftarrow{\mathbf{A}}^{(v)}\\
\underrightarrow{\mathbf{A}}^{(v)}
\end{pmatrix} ,
\end{equation}
where we have defined the vectors $\underleftarrow{\underrightarrow{\tilde{\mathbf{A}}}}^{(m)}$ as the field amplitudes in the bulk media
multiplied by the corresponding phase factors $e^{-i \lambda^{(s,m)}_{\nu}d}$, i.e. the field amplitudes at the bulk/grating interface ($z=-d$). The grating transfer matrix
$\Theta$ is a  $2\times 2$ block matrix, with each of the blocks defined as 
\begin{subequations}
\begin{gather}
\Theta_{11}=\underleftarrow{\mathcal{Y}}^{(m)\dag}\cdot\mathbb{G}(d) \cdot\underleftarrow{\mathcal{Y}}^{(v)} ,\\
\Theta_{12}=\underleftarrow{\mathcal{Y}}^{(m)\dag}\cdot\mathbb{G}(d) \cdot\underrightarrow{\mathcal{Y}}^{(v)} ,\\
\Theta_{21}=\underrightarrow{\mathcal{Y}}^{(m)\dag}\cdot\mathbb{G}(d) \cdot\underleftarrow{\mathcal{Y}}^{(v)} ,\\
\Theta_{22}=\underrightarrow{\mathcal{Y}}^{(m)\dag}\cdot\mathbb{G}(d) \cdot\underrightarrow{\mathcal{Y}}^{(v)} ,
\end{gather}
\label{theta-matrices}
\end{subequations}
where the ``grating" operator $\mathbb{G}(d)$ is
\begin{align}
\mathbb{G}(d)
&= \underleftarrow{\mathcal{Y}}^{(g)}\cdot\underleftarrow{\mathcal{P}}^{(g)}(-d)\cdot\underleftarrow{\mathcal{Y}}^{(g)\dag} +
\underrightarrow{\mathcal{Y}}^{(g)}\cdot\underrightarrow{\mathcal{P}}^{(g)}(-d)\cdot\underrightarrow{\mathcal{Y}}^{(g)\dag} \nonumber \\
&= \sum_{\nu,s} |\mathbf{Y}^{(s)}[\lambda^{(s,g)}_{\nu}]\rangle\langle\overline{\mathbf{Y}}^{(s)}[\lambda^{(s,g)}_{\nu}]| e^{-\imath \lambda^{(s,g)}_{\nu} d}\nonumber \\
&\,+(\lambda^{(s,g)}_{\nu}\to-\lambda^{(s,g)}_{\nu})
\equiv \mathbb{G}^{(e)}(d) + \mathbb{G}^{(h)}(d) ,
\label{grating-propagator} 
\end{align}
Thus, the operator $\mathbb{G}(d)$ is a 
decomposition in function of the polarization as well as the right and left eigenvectors. The grating operator describes the propagation
of the electromagnetic field through the grating, and it is directly related to the Green tensor of the electromagnetic field in the modulated
region.

Once we have obtained the theta-matrices we can get immediately the scattering operators:
\begin{subequations}
\begin{eqnarray}
\underleftarrow{\mathcal{R}} &=&-\Theta^{-1}_{22}\cdot\Theta_{21} , \\
\underrightarrow{\mathcal{R}} &=&\Theta_{12}\cdot\Theta^{-1}_{22} , \\
\underleftarrow{\mathcal{T}} &=&\Theta_{11}-\Theta_{12}\cdot\Theta^{-1}_{22}\cdot\Theta_{21}, \\
\underrightarrow{\mathcal{T}} &=& \Theta^{-1}_{22} .
\end{eqnarray}
\label{scattOps}
\end{subequations}
The logic behind the previous expressions is simple: $\underleftarrow{\mathcal{R}}$ is the matrix that gives the field amplitude $\underrightarrow{\mathbf{A}}^{(v)}$ in terms of  $\underleftarrow{\mathbf{A}}^{(v)}$. 
Similarly, $\underleftarrow{\mathcal{T}}$ connects $\underleftarrow{\tilde{\mathbf{A}}}^{(m)}$ with $\underleftarrow{\mathbf{A}}^{(v)}$, etc.
As one can see, the derivation reflection and transmission operators requires the inversion of the matrix $\Theta_{22}$, which is called the pivotal matrix \cite{Li93}. This matrix contains important information about the scattering properties of the grating.
Indeed, the zeros of its determinant are connected with the resonances of the scattering operators. A simple check of this property can be found in the next section, where the resonances are the surface plasmons for a plane metal-dielectric interface.  Despite being formally simple, the inversion of this matrix may pose
numerical problems because the matrix is sparse, can be singular (ill conditioned),  and may lead to numerical instabilities.  We will see in the last section how one can skirt this problem when computing the Casimir energy. 
As a last remark let us notice that, from the properties of the eigenvalues and of the eigenfunctions, it follows immediately that all scattering operators are symmetric in $\alpha_{0}$ and $k_{y}$. This information will simplify the calculation of the Casimir interaction at the end of this paper. 


\subsection{Special case: planar interface}
To validate the previous approach and clarify how the actual calculation works, let us consider the simple case of a planar interface between two homogeneous media (``$m$'' and ``$v$''). In this case we should recover the expression for the Fresnel reflection amplitudes in the $e$ and $h$ polarization basis. 
For $d=0$ the operator $\mathbb{G}$ becomes the identity, simplifying the expression of  the theta-matrices \eqref{theta-matrices}, which become block matrices, with each block having a dimension $2\times 2$. Using the expressions for the eigenvalues \eqref{eigenvalues-homogeneous} and 
eigenfunctions \eqref{eigenfunctions-homogeneous} in the homogeneous regions, we obtain all the elements of the pivotal matrix:
\begin{widetext}

\begin{align}
&[\Theta^{(ee)}_{22}]_{\gamma\nu}=\frac{\sqrt{\mu^{(m)}\mu^{(v)}}}{2}\left(
\frac{1}{\mu^{(m)}}+\frac{1}{\mu^{(v)}}\frac{\lambda_{\nu}^{(m)}}{\lambda_{\nu}^{(v) }}\frac{[\lambda^{(v)}_{\nu}]^{2}+k_{z}^{2}}{[\lambda_{\nu}^{(m)}]^{2}+k_{z}^{2}}\right)\delta_{\gamma\nu}, \nonumber\\
&[\Theta^{(hh)}_{22}]_{\gamma\nu}= \frac{\sqrt{\epsilon^{(m)}\epsilon^{(v)}}}{2}
\left( \frac{1}{\epsilon^{(m)}}+\frac{1}{\epsilon^{(v)}}\frac{\lambda_{\nu}^{(m)}}{\lambda_{\nu}^{(v)}}\frac{[\lambda_{\nu}^{(v)}]^{2}+k_{z}^{2}}{[\lambda_{\nu}^{(m)}]^{2}+k_{z}^{2}}\right)\delta_{\gamma\nu}, \nonumber\\
&[\Theta^{(eh)}_{22}]_{\gamma\nu}=\frac{\sqrt{\mu^{(m)}\epsilon^{(v)}}}{2} \frac{k_{z}\alpha_{\nu}}{\mu^{(m)}\epsilon^{(v)}\omega\lambda_{\nu}^{(v)}}\left(
1-\frac{[\lambda_{\nu}^{(v)}]^{2}+k_{z}^{2}}{[\lambda_{\nu}^{(m)}]^{2}+k_{z}^{2}}\right)\delta_{\gamma\nu}, \nonumber\\
&[\Theta^{(he)}_{22}]_{\gamma\nu}=-\frac{\sqrt{\epsilon^{(m)}\mu^{(v)}}}{2}\frac{k_{z}\alpha_{\nu}}{\epsilon^{(m)}\mu^{(v)}\omega\lambda_{\nu}^{v}}\left(
1-\frac{[\lambda_{\nu}^{(v)}]^{2}+k_{z}^{2}}{[\lambda_{\nu}^{(m)}]^{2}+k_{z}^{2}}\right)\delta_{\gamma\nu} .
\end{align}
As before,  $\alpha_{\nu} = \alpha_0 + 2 \pi \nu/p$.
The other matrices can be immediately derived from the previous expression by accordingly changing the sign of $\lambda$. For example for $\Theta^{\gamma\nu}_{12}$, $\lambda^{m}\to-\lambda^{m}$ while for $\Theta^{\gamma\nu}_{21}$, $\lambda^{v}\to-\lambda^{v}$, etc.. This also means that all $\Theta$-matrices are block diagonal with each block being $2 \times 2$ and, therefore, the same occurs for the reflection operator. In the special case $k_{y}=0$ 
the $(e,h)$ polarization basis coincides with the usual transverse electric (TE) and transverse magnetic (TM) polarization basis. In this case
the blocks and therefore the reflection operators are diagonal and, as an example, we have
\begin{align}
\underleftarrow{\mathcal{R}}=-
\begin{pmatrix}
\ddots&&&\\
&\frac{\mu^{v}\lambda^{m}-\mu^{m}\lambda_{\nu}^{v}}{\mu^{v}\lambda_{\nu}^{m}+\mu^{m}\lambda_{\nu}^{v}}&0&\\
&0&\frac{\epsilon^{v}\lambda_{\nu}^{m}-\epsilon^{m}\lambda_{\nu}^{v}}{\epsilon^{v}\lambda_{\nu}^{m}+\mu^{m}\lambda_{\nu}^{v}}&\\
&&&\ddots
\end{pmatrix}
=
\begin{pmatrix}
\ddots&&&\\
&r^{\rm TE}_{k_{x}=\alpha_{\nu}, k_{y=0}}&0&\\
&0&r^{\rm TM}_{k_{x}=\alpha_{\nu},k_{y=0}}&\\
&&&\ddots
\end{pmatrix} ,
\end{align}
where $r^{\rm TE, TM}$ are the usual Fresnel reflection amplitudes. 
\end{widetext}


\section{Eigenvalues: analytics and numerics}

It should be clear from the previous section that the key element to calculate the scattering operators are the solutions of the transcendental equation. In order to study them both analytically and numerically, it is convenient
to define the variable $\eta \equiv \gamma_1^2$, write $\gamma_2^2=\eta + [\mu_{2}(\omega)\epsilon_{2}(\omega) -\mu_{1}(\omega)\epsilon_{1}(\omega)] \omega^2$, and re-write the transcendental equation
in terms of the variable $\eta$ as $\tilde{D}(\eta)=0$. The advantage of doing this is that, in contrast to  \eqref{TrEq}, this new equation does not depend on
$k_y$, thereby reducing the dimensionality of the space where the solutions are defined. Once we solve for $\eta$, we obtain the original eigenvalues
$\lambda$ using $\lambda^2 =\mu_{1}(\omega)\epsilon_{1}(\omega) \omega^2-(k_y^2+\eta)$.

In general, the solutions of the transcendental equation $\tilde{D}(\eta)=0$ must be searched for numerically. This task is complicated by the fact that, for real physical frequencies, they are complex numbers. However, since the Casimir free energy \eqref{freeEnergy} is given as a sum over the pure imaginary Matsubara frequencies, in this section we consider the solutions of the transcendental equation already at imaginary frequencies, namely
\begin{widetext}
\begin{eqnarray}
0= {\tilde D}^{(s)}(\eta) &=& -\cos(\alpha_{0} p) + \cos(p_{1} \sqrt{\eta}) \cos(p_{2} \sqrt{\eta - [\epsilon(i \xi) -1] \xi^2}) \nonumber \\
&& -
\frac{1}{2}
\left(\frac{ \sqrt{\eta - [\epsilon(i \xi) -1] \xi^2}   }{\sigma^{(s)}_{2}(i \xi) \sqrt{\eta}}+
\frac{\sigma^{(s)}_{2}(i \xi)  \sqrt{\eta} }{ \sqrt{\eta - [\epsilon(i \xi) -1] \xi^2} 
}\right)
\sin(p_{1}\sqrt{\eta}) \sin(p_{2} \sqrt{\eta - [\epsilon(i \xi) -1] \xi^2} )  ,
\label{transcental-imaginary}  
\end{eqnarray}
\end{widetext}
where, for simplicity, hereafter we  specialize to the case where one of the medium is vacuum ($\epsilon_{1},\mu_{1}=1$) and the other has no magnetic activity ($\epsilon_{2}=\epsilon$, $\mu_{2}=1$). 
Our derivations and discussions below can be generalized to other grating configurations, where, for example, instead of vacuum we consider other materials, such as dielectrics or semiconductors.
We also recall that in the previous expression, $\sigma^{(e)}_{2}(i \xi)=1$ and $\sigma^{(h)}_{2}(i \xi) = \epsilon(i \xi)$.
One can analytically show that on the imaginary frequency axis $\omega=i \xi$, the solutions for $\eta=\eta[\xi,\alpha_{0}]$ are non-negative, real numbers. 
The eigenvalues are then purely imaginary quanties, $\lambda = \pm i \sqrt{ \xi^2 + k_y^2 + \eta }$.
Eq.(\ref{transcental-imaginary}) is the main equation in this work, that we shall study in detail below.


\subsection{Drude and plasma models for metallic gratings}
\label{DPeigen}

Depending on the range of frequency, the dielectric model, and the polarization it is possible to find approximate analytical expressions for the eigenvalues in some limiting cases. 
For simplicity, we will consider here only two model dielectric functions of metals, namely the Drude ($\epsilon_D$) and plasma ($\epsilon_p$) permittivities:
\begin{equation}
\epsilon_D(i\xi) = 1 + \frac{\omega_{p}^2}{\xi(\xi+\gamma)} ,\quad \epsilon_p(i\xi) = 1 + \frac{\omega_{p}^{2}}{\xi^2} ,
\end{equation}
where $\omega_{p}$ is the plasma frequency and $\gamma$ the dissipation rate.
In the following we study the high and low frequency behavior of the eigenvalues for both permittivity models.

As we stated in Section II, for $\xi \gg \omega_{p}$ the ultraviolet transparency of metals implies that the Drude and plasma models share the same set of eigenvalues, independent of polarization. The large eigenvalues ($\eta\gg \omega_{p}^{2}$) have the form
\begin{equation}
\eta(\xi \rightarrow \infty) = (\alpha_0+2 \pi \nu/p)^2,
\label{eigen-high}
\end{equation}
with $\nu \in \mathbb{Z}$, while for $\eta\lesssim \omega_{p}^{2}$ their values must be found numerically. 

At low frequencies the eigenvalues depend on polarization, and they are different for the Drude and plasma
models. We will call ``low frequency"  different regions for each of these models: for the Drude
mode it corresponds to $\xi\ll \gamma$, while for the plasma model to $\xi \ll \omega_{p}$.  In the region $\gamma\ll \xi \ll \omega_{p}$ the solutions for the two polarizations behave differently, but the plasma and the Drude model give similar expressions. In the region $\xi \ll \gamma$, absent in the plasma model, the Drude model describes a regime where the electromagnetic field undergoes a diffusive dynamics \cite{Jackson75,Intravaia09}. We now consider the two polarizations separately.


\subsubsection{$s=h$ polarization}

\begin{figure}
\begin{center}
\includegraphics[width=3.57cm]{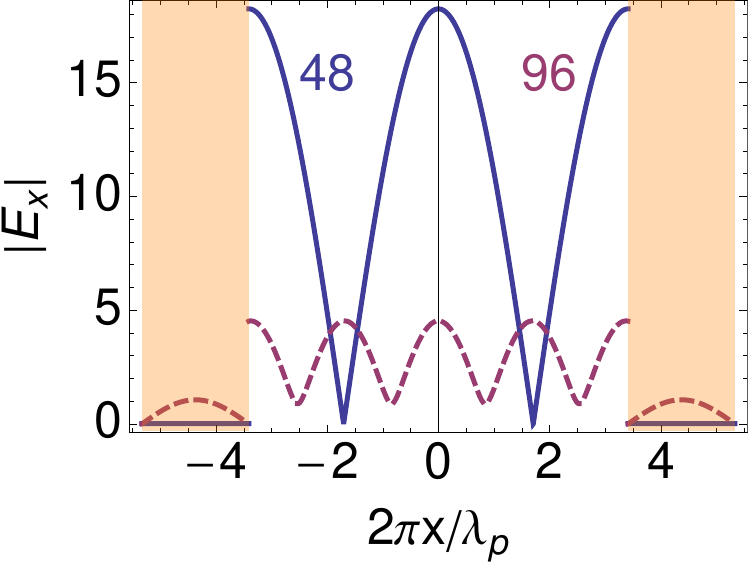}\hspace{1mm}
\includegraphics[width=3.8cm]{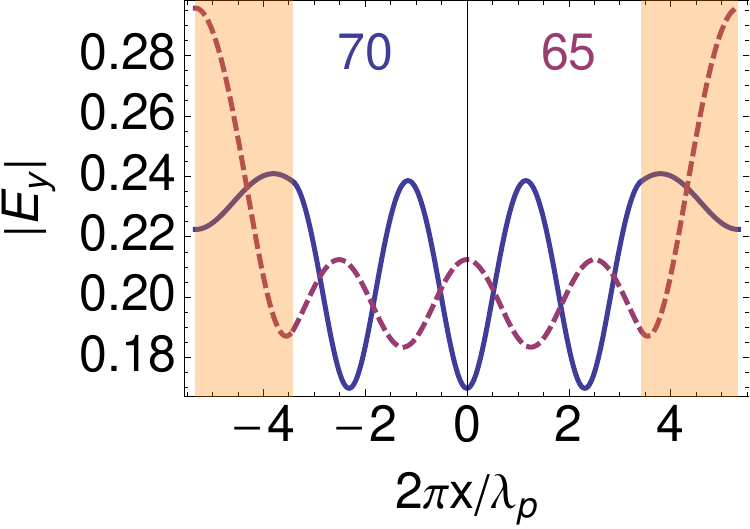}
\includegraphics[width=3.6cm]{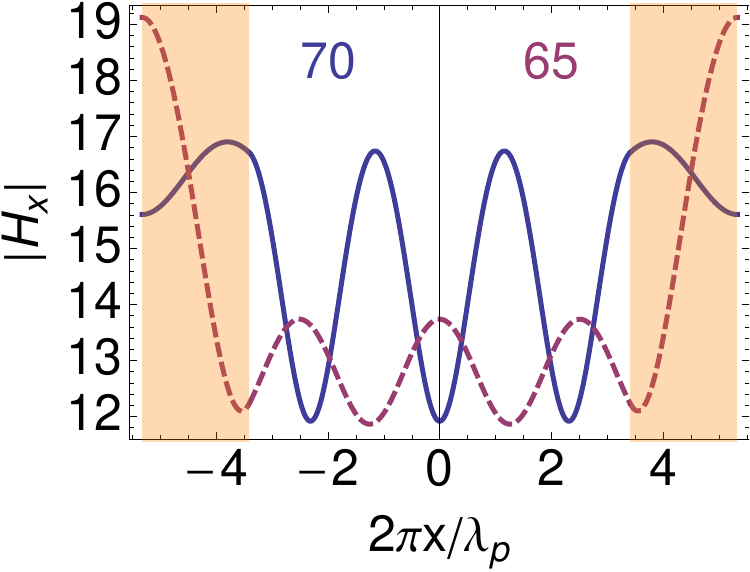}\hspace{1mm}
\includegraphics[width=3.7cm]{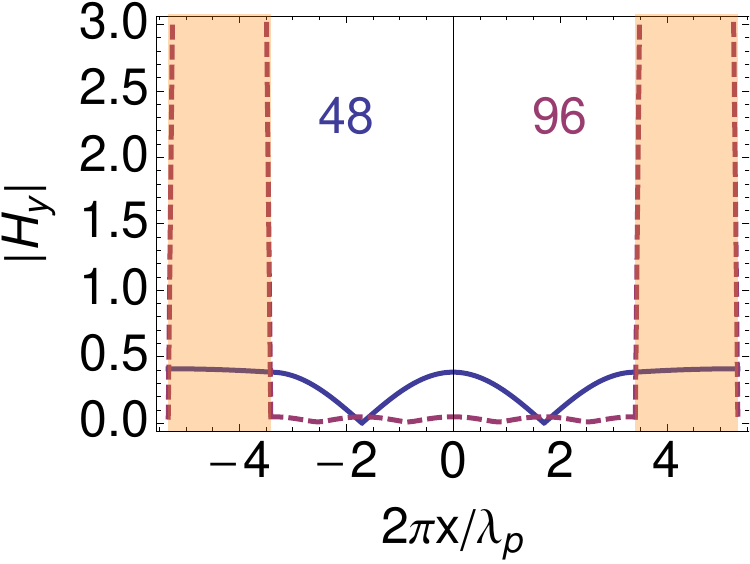}
\end{center}
\vspace{-0.8cm}
\caption{Spatial structure of the electromagnetic modes in the grating region in units of the plasma wavelength $\lambda_p=2 \pi/\omega_p$,
for  $k_{y}=0$, $\alpha_0=0.2 \pi/p$ for the second Matsubara frequency $\xi= 4 \pi k_{B}T/\hbar$ at$T=300$ K.
The curves represent the modes with $\nu=2$, corresponding to the two seeds in Eqs. \eqref{solh1} and \eqref{solh2}, and to the two seeds 
($\pm$ solutions) in \eqref{sole}. Our choice $k_y=0$ implies that the $e$ and $h$ polarizations decouple, and that $E_{x}$ and $H_{y}$ depend only on the $h$-polarization, while $E_{y}$ and $H_{x}$ only on the $e$-polarization.
For the $h$-polarization two categories of modes exist: the first mainly vibrate within the grooves, and the second mainly within the teeth (this is particularly clear for $H_{y}$). In agreement with Maxwell equations, the component of the electric field along the modulation direction is discontinuous at the groove walls while the remaining ones are all continuous. The numerical values inside the plots indicate the effective refractive index at imaginary frequency, $n_{\rm eff}(i \xi) \equiv k_{z}/\xi$, for the corresponding mode (the value on the left corresponds to the full line mode, the one on the right to the dashed line mode).
The Drude and grating parameters are the same as in the previous figure. 
}
\label{ModePlot}
\end{figure}

In the limit $\xi\ll \omega_{p}$, assuming that $\eta[\xi,\alpha_{0}]$ is constant or goes to zero slower than $[\epsilon(i\xi)-1]\xi^{2}$, the $s=h$ value of the term in  the big parentheses in the second line of 
Eq.(\ref{transcental-imaginary}) is much larger than one. Then one has to look for solutions of
$\sin(p_{1}\sqrt{\eta}) \sin(p_{2}\sqrt{\eta-[\epsilon(i \xi)-1] \xi^2})=0$. Two sets of solutions are possible: the first
\begin{equation}
\eta^{(h)}_{1,\nu}(\xi\ll \omega_{p}) = \left(\frac{\nu \pi}{p_{1}}\right)^{2} ,
\label{solh1}
\end{equation}
($\nu\in \mathbb{Z}$, and $\nu \neq 0$) does not depend on the permittivity model  and describes modes vibrating within the grooves (see fig. \ref{ModePlot}); the second
one is given by
\begin{equation} 
\eta^{(h)}_{2,\nu}(\xi\ll \omega_{p}) =
\begin{cases}
\left(\frac{\nu \pi}{p_{2}}\right)^{2} +\omega_{p}^{2} & {\rm (plasma)} \\
\left(\frac{\nu \pi}{p_{2}}\right)^{2} + \frac{\xi \omega_p^2}{\xi +\gamma} &  {\rm (Drude)},
\end{cases}
\label{solh2}
\end{equation}
and describes modes vibrating inside the teeth (see fig. \ref{ModePlot}). The difference between the two dielectric models is evident in the limit $\xi \ll \gamma$.
For the plasma model all the solutions are always distinct. On the contrary, for the Drude model degeneracies are possible:  for certain frequencies $\xi$ there are values of $\nu$ that make the eigenvalues of Eq.\eqref{solh1} identical to the ones of Eq.\eqref{solh2}, and in this case an alternative approach must be used to search for the solutions (see the end of this Section). 

For $\eta[\xi,\alpha_{0}]$ going to zero faster than $[\epsilon(i\xi)-1]\xi^{2}$ for $\xi \to 0$, one can no longer neglect the terms in the first line Eq.(\ref{transcental-imaginary}). In this case, for $ \xi\ll \omega_{p}$
in the plasma model one can approximate 
$\eta-[\epsilon(\imath\xi)-1] \xi^2 \approx -\omega_{p}^{2}$, and expand up to the second order in $\eta$ the terms $\cos(p_{1}\sqrt{\eta})$ and  $\sin(p_{1}\sqrt{\eta})$.  Solving the resulting equation one gets
for the smallest eigenvalue
\begin{equation}
\eta^{(h)}_{\nu=0,{\rm plasma}}
\approx 2\xi^{2}\frac{\cosh(p_{2}\omega_p)-\cos(\alpha_{0}p)}{\omega_p p_{1}\sinh(p_{2}\omega_p)}.
\label{0plasma}
\end{equation}
which describes a mode resulting from the coupling of surface plasmons living on the walls of the grooves.
For the Drude model $\eta-[\epsilon(\imath\xi)-1] \xi^2$ goes also to zero for $\xi \to 0$. Expanding to the second order in $\eta$ the corresponding trigonometric functions, and solving for $\eta$ one gets
\begin{equation}
\eta^{(h)}_{\nu=0, {\rm Drude}}\approx 
2\xi\gamma\left\{
\frac{\left[ 1-\cos(\alpha_{0} p)\right]}{\omega_{p}^{2}p_{1}p_{2}}+\frac{1}{2}\frac{p_{2}}{p_{1}}\frac{\xi}{\gamma}\right\}.
\label{0Drude}
\end{equation}
Hence,  $\eta^{(h)}_{0, {\rm plasma}}$ goes quadratically to zero with the frequency,
while the corresponding power law $\eta^{(h)}_{0, {\rm Drude}}$
strongly depends on the value of $\alpha_{0}$. 
This last feature will be relevant in the numerical evaluations below, in particular in the calculation of the zero frequency limit of the reflection operators. 


\subsubsection{$s=e$ polarization}

Let us consider now the low frequency behavior of the eigenvalues in the case of $e$-polarization. 
For the plasma model one can see that Eq.(\ref{transcental-imaginary}) does no longer depend on the frequency, and in consequence the corresponding eigenvalues are frequency-independent and
coincide with their high-frequency limit. The eigenvalues must be found numerically, the large ones being
approximately equal to \eqref{eigen-high}.
For the Drude model Eq.(\ref{transcental-imaginary}) becomes identical to the one for vacuum. The solutions are then
\begin{equation}
\eta^{(e)}_{\rm Drude} (\xi \ll \gamma)=
\begin{cases}
\alpha_0^2 & {\rm for} \; \nu=0 \\
\left(\pm \alpha_{0}+\frac{2\pi \nu}{p}\right)^{2} & {\rm for} \; \nu \neq 0 .
\end{cases}
\label{sole}
\end{equation}
In this case degeneracies  happen at the center ($\alpha_{0}=0$) and at the border ($\alpha_{0}=\pi/p$) of the Brillouin zone \cite{Suratteau83} 
(see Section \ref{CasimirInteraction} for the impact of degeneracies on the calculation of the Casimir interaction).
Expanding the transcendental equation \eqref{transcental-imaginary}
to second order in $\eta$ around the solutions \eqref{sole}, and solving for $\eta$ one gets 
\begin{widetext}
\begin{equation}
\eta^{(e)}_{{\rm Drude}}= 
\begin{cases}
\left. \alpha_{0}^{2} -\frac{2\tilde{D}^{(e)}(\eta) }{\partial_{\eta}\tilde{D}^{(e)}(\eta)-\sqrt{[\partial_{\eta}\tilde{D}^{(e)}(\eta)]^{2}-4\tilde{D}_{\;}^{(e)}(\eta)\partial^{2}_{\eta}\tilde{D}^{(e)} (\eta)}} \right|_{\eta=\alpha_{0}^{2}} & {\rm for} \; \nu=0 \\
\left. \left(\pm \alpha_{0}+\frac{2\pi \nu}{p}\right)^{2} -\frac{2\tilde{D}^{(e)}(\eta)}{\partial_{\eta}\tilde{D}^{(e)}(\eta)\mp\sqrt{[\partial_{\eta}\tilde{D}^{(e)}(\eta)]^{2}-4\tilde{D}_{\;}^{(e)}(\eta)\partial^{2}_{\eta}\tilde{D}_{\eta^{2}}^{(e)}(\eta)}} 
\right|_{\eta=\left(\pm \alpha_{0}+2\pi \nu/p \right)^{2}} & {\rm for} \; \nu \neq 0 .
\end{cases}
\end{equation}
\end{widetext}
It is possible to show that in the limit $\alpha_{0}\to 0$
\begin{equation}
\eta^{(e)}_{\nu=0, {\rm Drude} }\approx\alpha_{0}^{2}+\frac{p_{2}}{p}\frac{\xi}{\gamma}.
\end{equation}
%


\begin{figure}
\begin{center}
\includegraphics[width=8.cm]{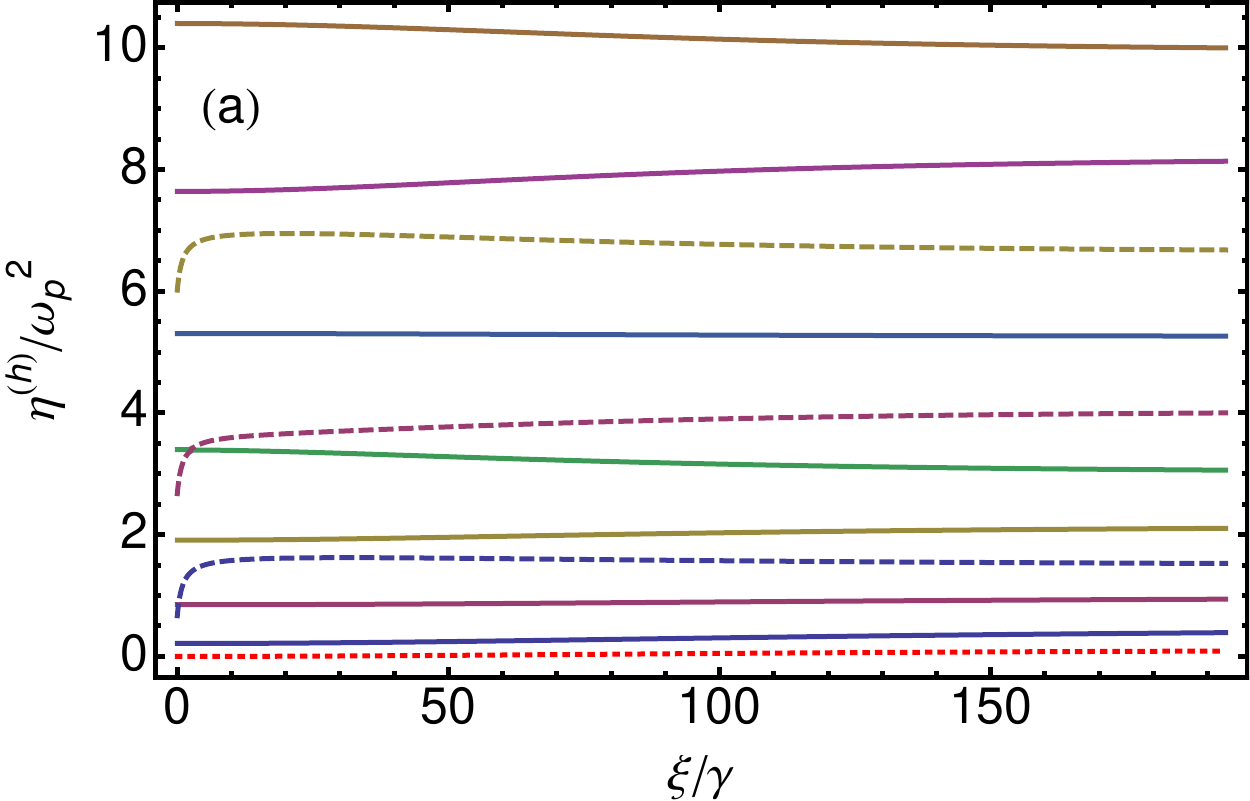}\vspace{1cm}
\includegraphics[width=8.cm]{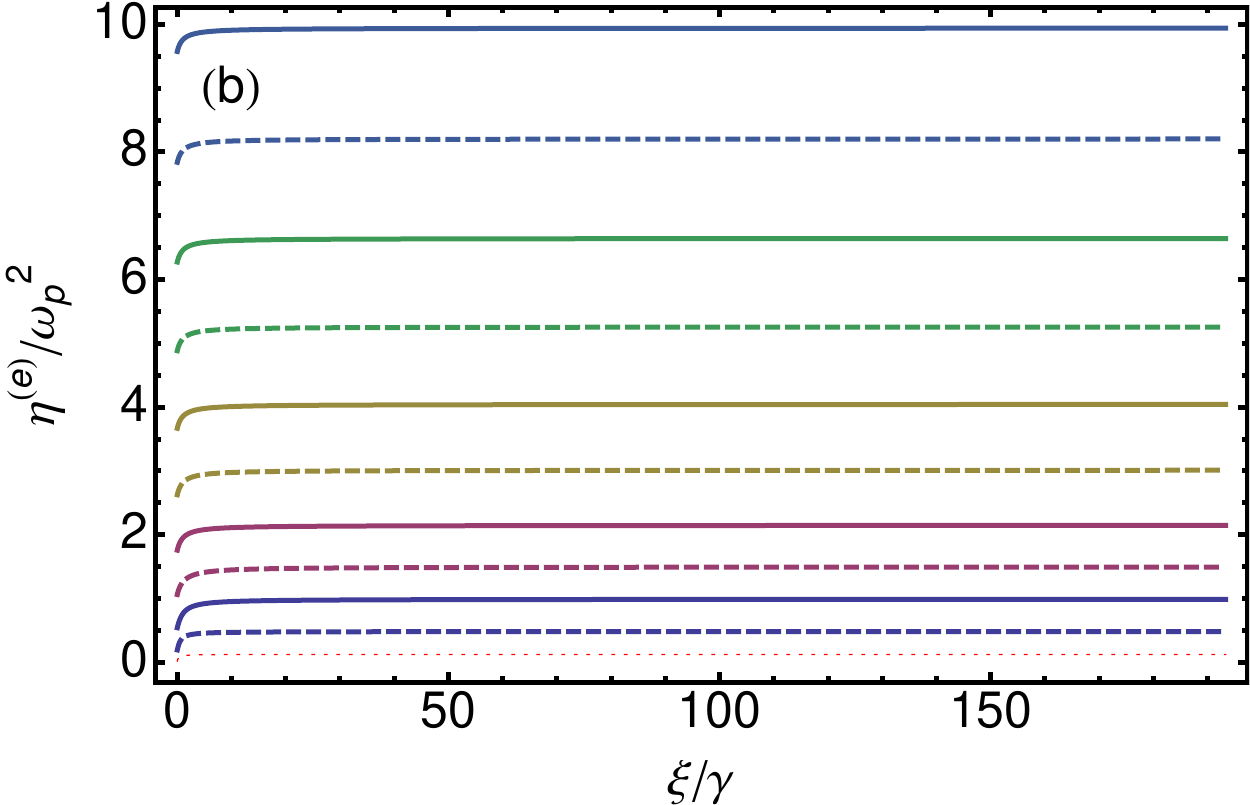}
\caption{Numerical solution of the transcendental equation (\ref{transcental-imaginary})
for the Drude model. Only the lowest eleven eigenvalues are shown. (a) $s=h$ polarization. The dotted line corresponds to the smallest eigenvalue $\eta^{(h)}_{\nu=0}$. 
Solid lines are the eigenvalues obtained using as seeds the expression \eqref{solh1} [$\nu$ going from 1 (bottom curve)
to 4 (top curve)]. Dashed lines  are the eigenvalues obtained using as seeds the expression\eqref{solh2} 
[$\nu$ going from 1 (bottom curve) to 2 (top curve)].
(b)  $s=e$ polarization.
The dotted line corresponds to the smallest eigenvalue $\eta^{(e)}_{\nu=0}$. 
Solid and dashed lines are the eigenvalues obtained using as seeds the expressions \eqref{sole} 
for the two possible signs [$\nu$ going from 1 (bottom curve)
to 4 (top curve)]. Dashed lines  are the eigenvalues obtained using as seeds the expression \eqref{solh2} 
[$\nu$ going from 1 (bottom curve) to 3 (top curve)].
Parameters are $p_{1}=160$ nm, $p_{2}=90$ nm, and $\alpha_{0}=0.5 \pi/p$. The optical parameters
chosen for these plots are $\omega_{p}=8.39$ eV, $\gamma=0.043$ eV; the general structure of the curves remains
unchanged for other choices of Drude parameters.  
}
\label{PlotEigen}
\end{center}
\end{figure}

\begin{figure}[t]
\includegraphics[width=8.cm]{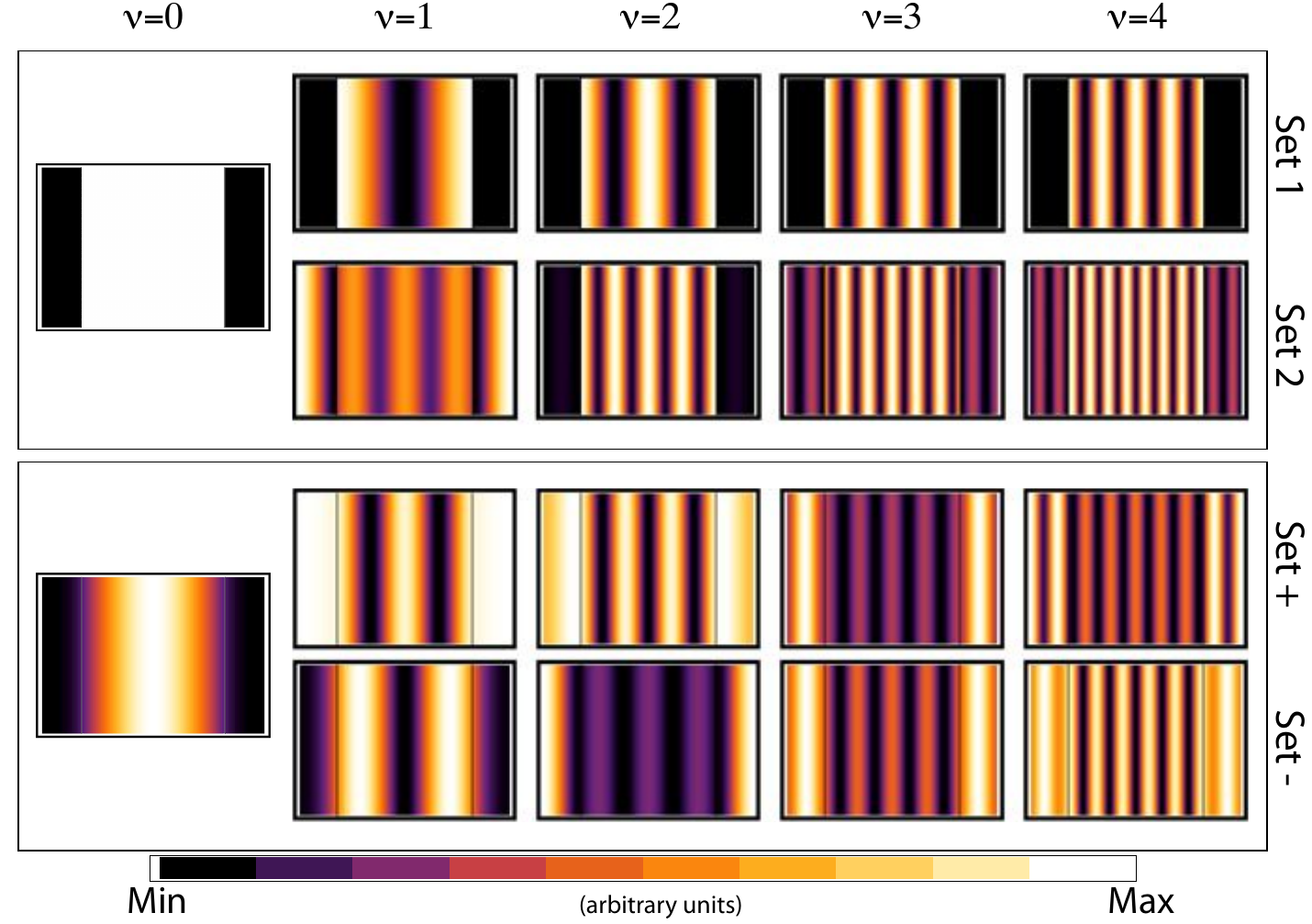}
\vspace{-0.4cm}
\caption{Density plot of the intensity of the electromagnetic field within the modulated region. Upper box corresponds to $|E_x|^2$ and lower box to
$|E_y|^2$.  In each of the plots the horizontal axis is the modulation $x-$direction and the vertical axis is the invariant $z-$ direction.
The parameters are the same of the previous figure and, therefore, $|E_{x}^{2}|$ depends only on the $h$-polarization (eqs.\eqref{solh1} and \eqref{solh2}) while $|E_{y}^{2}|$ depends only on the $e$-polarization (eq.\eqref{sole}). The first five modes for each polarization and for each set of eigenvalues are represented. 
There is only one zero mode per polarization (see discussion in the text). For the parameters chosen here the zero mode is almost constant for the $h$-polarization.
}
\label{Modes}
\end{figure}


\subsection{Numerical solution for eigenvalues}

We solved numerically the transcendental equation \eqref{transcental-imaginary} 
using Mathematica, employing as seeds for the roots of 
$\tilde{D}^{(s)}(\eta)=0$ the asymptotic analytical expressions for the eigenvalues described above. 
Hereafter we will focus 
on the results for the Drude model, postponing the results for the plasma model and their comparison for future work.
In figure \ref{PlotEigen} we show the numerical solutions for specific values of the geometrical parameters of the grating. As seen in the figure, all $s=e$ eigenvalues bend down at low frequencies, while
only the $s=h$ eigenvalues obtained from the seeds \eqref{solh2} show the same trend. This behavior is due to the
dissipative nature of the metal. 
For the $h$ polarization the eigenvalues obtained from the seeds \eqref{solh1} change
smoothly and they cross the other set of $h$ eigenvalues for some values of $\xi$, showing degeneracies. 
Both sets of eigenvalues of the $h-$polarization are almost insensitive to the value of $\alpha_{0}$, while this parameter becomes relevant at large imaginary frequency. In contrast, the eigenvalues of the $e$-polarization 
are very sensitive to the value of $\alpha_{0}$. In figs. \ref{ModePlot} and \ref{Modes} we plot the spatial profile of the eigenmodes corresponding to
the previously discussed eigenvalues.


\section{Details on the calculation of the matrix elements}

Now that we have described the calculation of the eigenvalues and the eigenvectors, let us proceed to the computation
of the theta-matrices \eqref{theta-matrices}. All these matrices involved in the calculation of the Casimir free energy have a similar form, namely a collection of $2\times 2$ blocks with elements coupling the two polarizations:
\begin{equation}
[\Theta_{ij}]^{(ss')}_{\gamma\nu}=
\langle\overline{\mathbf{Y}}^{(s,m)}[(-1)^{i}\lambda_{\gamma}]|\mathbb{G}(d)| \mathbf{Y}^{(s',v)}[(-1)^{j}\lambda_{\nu}]\rangle .
\end{equation}
From the expression for the grating operator \eqref{grating-propagator} and of the eigenvectors, it follows that one of the key elements of our approach is the calculation of the overlap between the eigenvectors describing the field in the grating region with the eigenvectors characterizing the field in the two homogeneous regions, namely
$\langle\overline{\mathbf{Y}}^{(s)}[\lambda^{(s,m)}_{\nu}] | \mathbf{Y}^{(s')}[\lambda^{(s',g)}_{\nu'}]\rangle$ and
$\langle\overline{\mathbf{Y}}^{(s)}[\lambda^{(s,g)}_{\nu}] | \mathbf{Y}^{(s')}[\lambda^{(s',v)}_{\nu'}]\rangle$, and 
eventually the explicit calculation of the following integrals
\begin{equation}
\int^{\frac{p}{2}}_{-\frac{p}{2}}\mathrm{d}x\ \frac{\mathcal{U}^{(s)}[x,\lambda]}{\sigma^{(s)}(x)}e^{-\imath\alpha x},\quad \int^{\frac{p}{2}}_{-\frac{p}{2}}\mathrm{d}x\ \frac{\partial_{x}\mathcal{U}^{(s)}[x,\lambda]}{\sigma^{(s)}(x)}e^{-\imath\alpha x}.
\end{equation}
It is interesting to note that since $\mathcal{U}^{(s)}[x,\lambda]$, defined in \eqref{definition-U}, is a combination of trigonometric functions oscillating with frequencies $\gamma_1$ and $\gamma_{2}$, the above integrals are large when
$\alpha=\pm\gamma_{i}\quad(i=1,2)$.
Physically speaking this relation describes the $x$-component momentum matching between the electromagnetic wave coming from the homogeneous regions and the wave propagating in the grating region. The above integrals can be done analytically, but the resulting expressions are long and cumbersome,
so we do not report them here.

It is interesting to consider some special cases. For instance, for $k_{y} = 0$ the scalar products between vectors with different polarizations vanish. As a consequence the polarizations decouple and one can show that the blocks of the theta-matrices become diagonal.  Under a transformation that generates an even number of permutations of rows and  columns, the transfer matrix can be written in a block diagonal form as 
\begin{equation}
\Theta(k_{y}=0)=
\begin{pmatrix}
\Theta^{(ee)}&0\\
0&\Theta^{(hh)}
\end{pmatrix} .
\end{equation} 
From Eqs.\eqref{scattOps} it immediately follows that all scattering operators are $2\times2$ block diagonal. 
In the case where the Drude model is used to describe the optical properties of the metallic grating, some interesting information can be obtained for the reflection operator in the limit $\xi\to 0$ for the $e$-polarization. In this limit
$\mathcal{U}^{(e)}=\mathcal{U}_{\rm hom}^{(e)}$ because the eigenvalues \eqref{sole} are identical to the ones of vacuum, 
i.e., the grating modes match the ones of the vacuum region,  and therefore the electromagnetic field effectively does not see the grating modulation. The properties of the $e$-polarization allow to directly connect this result to the Bohr-van Leeuwen theorem \cite{Leeuwen21,Bimonte09}.

Decomposing the operator $\mathbb{G}$ over the two polarizations, in the limit $\xi\to 0$ and arbitrary $k_y$ we can write
\begin{widetext}
\begin{align}
\Theta_{\gamma\nu}(\xi= 0)
&=
\begin{pmatrix}
\langle\overline{\mathbf{Y}}^{(e)}[\lambda^{(e,m)}_{\gamma}]| \mathbb{G}^{(h)} | \mathbf{Y}^{(e)}[\lambda^{(e,v)}_{\nu}]\rangle
&\langle\overline{\mathbf{Y}}^{(e)}[\lambda^{(e,m)}_{\gamma}] | \mathbb{G}^{(h)} |\mathbf{Y}^{(h)}[\lambda^{(h,v}_{\nu}]\rangle\\
\langle\overline{\mathbf{Y}}^{(h)}[\lambda^{(h,m)}_{\gamma}] | \mathbb{G}^{(h)} |\mathbf{Y}^{(e)}[\lambda^{(e,v)}_{\nu}]\rangle
&\langle\overline{\mathbf{Y}}^{(h)}[\lambda^{(h,m)}_{\gamma}] | \mathbb{G}^{(h)} | \mathbf{Y}^{(h)}[\lambda^{(h,v)}_{\nu}]\rangle
\end{pmatrix}
+
\begin{pmatrix}
\delta_{\lambda_{\gamma},\lambda_{\nu}}e^{-{\rm sign}[\lambda_{\gamma}]d\sqrt{k^{2}_{z}+\alpha^{2}_{\nu}}}&0\\
0&0
\end{pmatrix} ,
\end{align} 
\end{widetext}
where the first term corresponds to the $h$ part of the operator $\mathbb{G}$ and the second one to the $e$ part.
Here $\lambda_{\gamma}$ and $\lambda_{\nu}$ can be positive and negative.

If we now consider in addition the limit $k_{y}\to0$,  we can deduce the following properties for the reflection matrix. Since $\Theta_{ij}^{(ee)}=\Theta_{ij}^{(eh)}=\Theta_{ij}^{(he)}=0$ for $i\not=j$ we immediately have that
\begin{gather}
\underleftarrow{\mathcal{R}}^{(ee)}(\xi=0,k_{y}=0)=0 , \nonumber\\
\underleftarrow{\mathcal{R}}^{(eh)}(\xi=0,k_{y}=0)=
\underleftarrow{\mathcal{R}}^{(e,h)}(\xi=0,k_{y}=0)=0.
\label{zero-limit-1}
\end{gather}
The only part of the reflection operator which does not vanish is connected with the $h$-polarization:
\begin{gather}
\underleftarrow{\mathcal{R}}^{(hh)}(\xi=0,k_{y}=0)=-[\Theta_{22}^{(hh)}]^{-1}\Theta_{21}^{(hh)}. 
\end{gather}
A rather lengthy calculation also allows us to obtain some properties of the previous operator matrix elements. We only report
the most relevant one for the first Brillouin zone, i.e. in the limit $\alpha_0 p/\pi \ll 1$
\begin{equation}
\underleftarrow{\mathcal{R}}_{00}^{(hh)}(\xi=0,k_{y}=0) -  1 \propto  -\alpha_0.
\label{zero-limit-2}
\end{equation}

\begin{figure}[t]
\includegraphics[width=8.5cm]{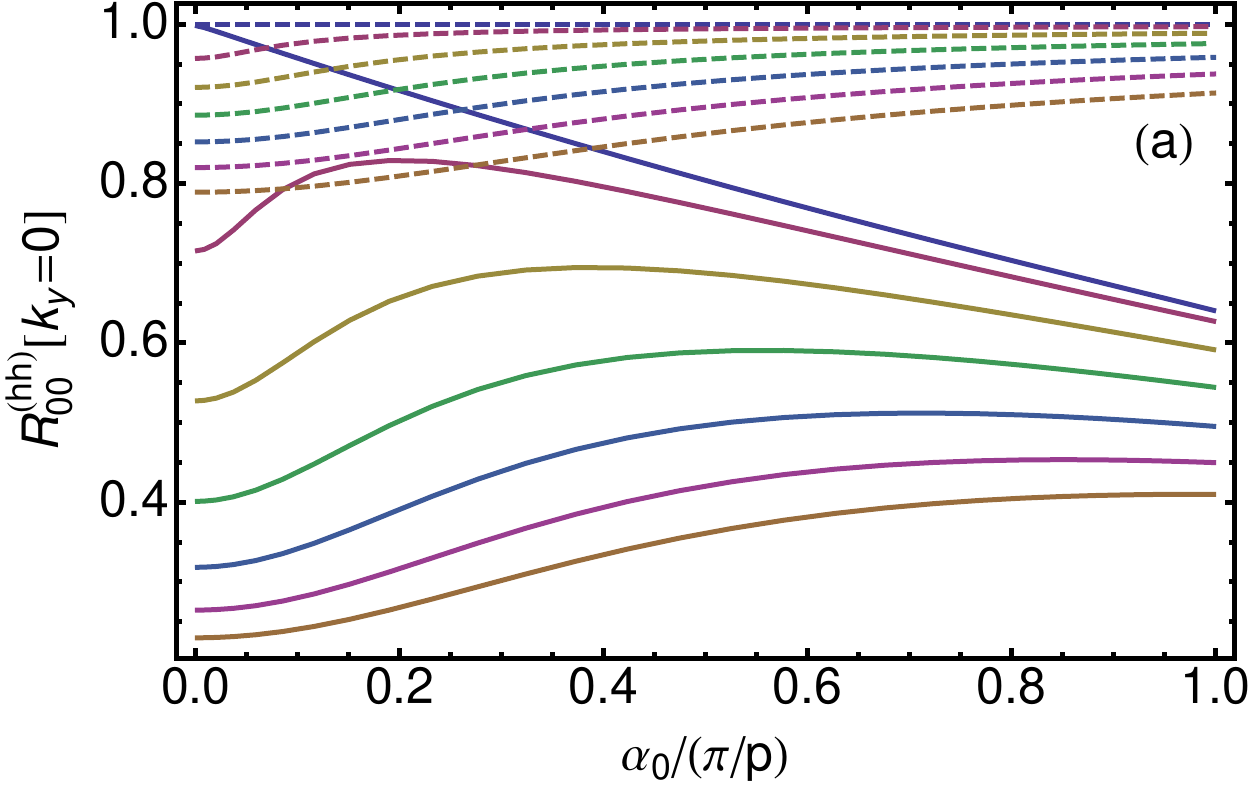}\vspace{5mm}
\includegraphics[width=8.5cm]{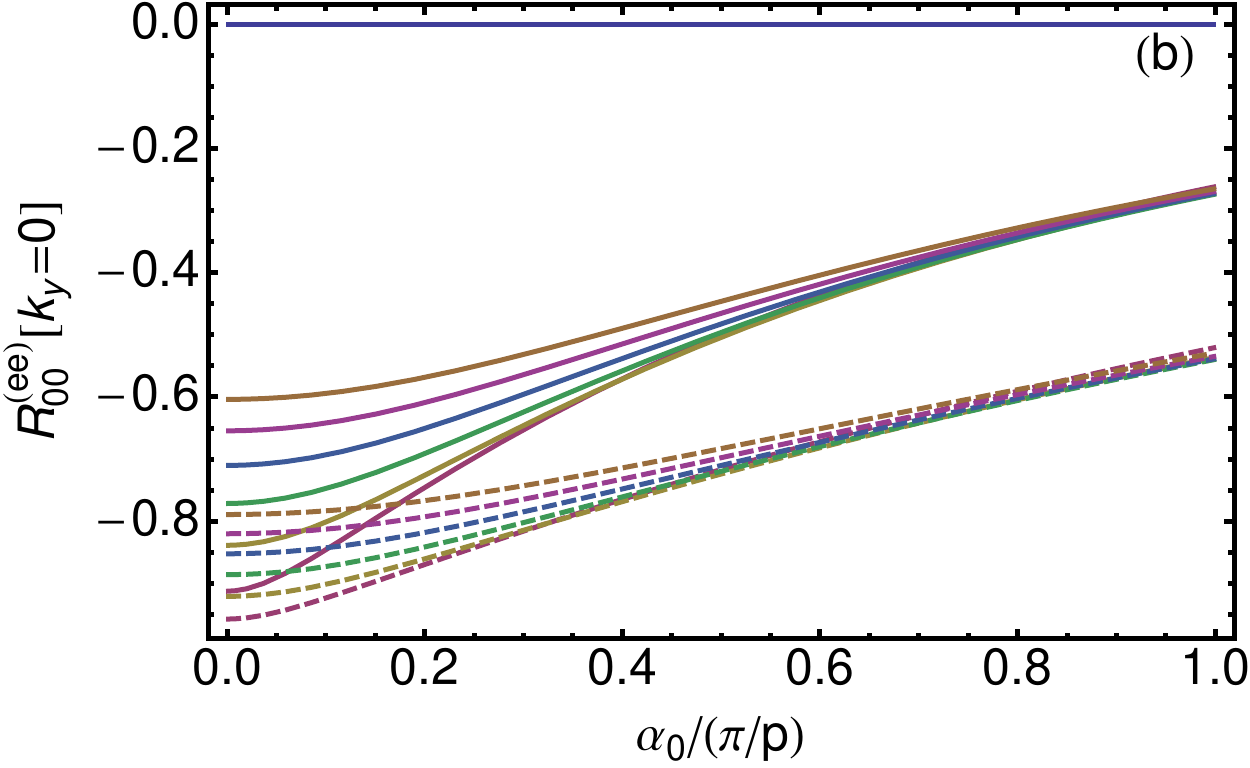}
\caption{Matrix elements of the reflection operator corresponding to the zeroth order reflection.  The full lines are the result for the grating for the first seven Matsubara frequencies. The dashed lines show the equivalent matrix elements for the flat surface (Fresnel coefficients). 
(a): 
zeroth order reflection coefficients for the $h$-polarization. The first seven  Matsubara frequencies are shown (zeroth to the sixth from the top to the bottom).
(b):
zeroth order reflection coefficients for the $e$-polarization. Once again, the first seven  Matsubara frequencies are shown. The zeroth frequency term is zero, while the other ones (first to the sixth from the bottom to the top) decrease in absolute value.
The depth of the grating is $d=216$ nm, and the remaining parameters are the same as in fig. \ref{PlotEigen}.
}
\label{Reflections}
\end{figure}

The solid lines in figure \ref{Reflections} are  the numerically computed matrix elements of the reflection operator of the metallic grating corresponding to the zeroth order reflection ($\gamma=\nu=0$) for the first
seven Matsubara frequencies, shown only for $k_y=0$. The Drude model was used with $\omega_{p}=8.39$ eV, $\gamma=0.043$ eV, while the grating geometry is:  $p_{1}=160$ nm,
$p_{2}=90$ nm, and $d=216$ nm.
The behavior of these reflection amplitudes is in agreement with the predictions made above. The dashed lines show the corresponding matrix elements  or a flat  metallic surface (Fresnel coefficients), using the same optical parameters.
From the figure it is clear that the grating is less specularly reflecting than a flat surface. 
At large wavevectors the grating reflection amplitudes behave differently with respect to the plane surface: while for the plane they asymptotically reach a horizontal line, for the grating they have a finite negative slope.
However, at small wavevectors the behavior of
the reflection amplitudes for the grating and for the flat surface is similar (except for the zeroth Matsubara $hh$ reflection
amplitude). This suggest that in this limit the grating may be described using an effective medium approximation, in which
the reflection matrices of the grating are approximated by Fresnel coefficients for an homogeneous planar interface with
an effective permittivity $\epsilon_{\rm eff}(\omega)$.
A fit of the numerical results for the grating in figure \ref{Reflections} for the $h-$ and $e$-polarization to Fresnel coefficients gives an effective Drude permittivity with a reduction of the plasma frequency of about 7.8 times for $h$ polarization and 2.2 times for the $e$ polarization. The effective dissipation rate decreases more for the $e$-polarization than for the $h$-one (1.4 times against 1.2).

Finally, let us emphasize that our method allows us to deal with the zero Matsubara frequency ($\xi_{l=0}=0$) analytically, without resorting to any limiting
procedure, such as approximating $\xi_{l=0}$ by a large wavelength mode (as used in, for example, \cite{Davids10}). It also avoids problems related to the Gibbs phenomenon which complicates the calculation, especially for metallic structures.  This phenomenon refers to the oscillations that occur when a piecewise discontinuous function, such as our permittivity $\epsilon(x;\omega)$, is approximated by a finite Fourier series. Its impact increases with the magnitude of the discontinuity and, therefore, becomes a more serious issue at low frequencies. Instead, the method described in this paper deals
with such a discontinuity exactly, eliminating \textit{de facto} all problems relate with a Fourier decomposition of the permittivity profile.


\section{The Casimir Interaction}
\label{CasimirInteraction}
In this section we use our quasi-analytical modal approach to write down the Casimir free energy \eqref{freeEnergy} between two vacuum-separated,
lamellar gratings facing each other. We discuss how to generalize the formalism to non-lamellar gratings and multilayered periodic structures. Finally, 
 we numerically compute the Casimir  pressure between a flat gold plate parallel
to a gold grating, and discuss the asymptotic behaviors at large and small distances.

\subsection{Two lamellar gratings}

Let us consider two vacuum-separated lamellar 1D gratings.  The Casimir pressure between them can be obtained by taking the $a$ derivative
of the Casimir free energy \eqref{freeEnergy},
\begin{align}
P(a)&=-\frac{4}{\beta} \sum^{\infty'} _{l=0}  
\int_{0}^{\infty}\hspace{-0.3cm}dk_y \int_{0}^{\pi/p} \hspace{-0.3cm}d\alpha_0  \nonumber \\
&\times
\partial_{a}\log {\rm det} \left[1- \underleftarrow{\mathcal{R}}^{L} \cdot \underrightarrow{\mathcal{P}}^{(v)}(a) \cdot \underrightarrow{\mathcal{R}}^{R} 
\cdot \underrightarrow{\mathcal{P}}^{(v)}(a) \right] ,
\label{pressure}
\end{align}
where we have already performed the trace over the spatial degrees of freedom and used the parity properties of the reflection operators
\begin{equation}
\underleftarrow{\mathcal{R}}^{L}=
-[\Theta^{L}_{22}]^{-1}\cdot\Theta^{L}_{21},\quad \underrightarrow{\mathcal{R}}^{R}=\Theta^{R}_{12}\cdot[\Theta^{R}_{22}]^{-1} ,
\label{reflection-for-energy}
\end{equation}
and of $\underrightarrow{\mathcal{P}}^{(v)}(a)$. 
As we discussed above, the calculation of the reflection matrices requires the inversion of the pivotal matrix,  which can be an expensive and not accurate numerical operation. It is however possible to avoid this inversion and derive the Casimir free energy. 
Indeed, using \eqref{reflection-for-energy} in \eqref{pressure} we can write
\begin{align}
&\log {\rm det} \left[1- \underleftarrow{\mathcal{R}}^{L} \cdot \underrightarrow{\mathcal{P}}^{(v)} \cdot \underrightarrow{\mathcal{R}}^{R} 
\cdot \underrightarrow{\mathcal{P}}^{(v)} \right]
=\nonumber\\
&\log\frac{\mathrm{det}\left[\Theta^{L}_{21}\cdot\underleftarrow{\mathcal{P}}^{(v)}(-a)\cdot\Theta^{R}_{12}+\Theta^{L}_{22}\cdot\underrightarrow{\mathcal{P}}^{(v)}(-a)\cdot\Theta^{R}_{22}\right]}
{\mathrm{det}\left[\Theta^{L}_{22}\cdot\underrightarrow{\mathcal{P}^{(v)}}(-a)\cdot\Theta^{R}_{22}\right]}, 
\label{CasOp2}
\end{align} 
where the different theta-matrices for the left ($L$) and right ($R$) gratings can be obtained from \eqref{theta-matrices}, namely
\begin{eqnarray}
\Theta^{L}_{21} &=& \underrightarrow{\mathcal{Y}}^{(m)\dag}\cdot \mathbb{G}^{L}(d_L)\cdot\underleftarrow{\mathcal{Y}}^{(v)} , \nonumber \\
\Theta^{L}_{22} &=& \underrightarrow{\mathcal{Y}}^{(m)\dag}\cdot \mathbb{G}^{L}(d_L) \cdot\underrightarrow{\mathcal{Y}}^{(v)} , \nonumber \\ 
\Theta^{R}_{12} &=& \underleftarrow{\mathcal{Y}}^{(v)\dag}\cdot \mathbb{G}^{R}(d_R) \cdot\underrightarrow{\mathcal{Y}}^{m)} , \nonumber \\
\Theta^{R}_{22} &=&  \underrightarrow{\mathcal{Y}}^{(v)\dag}\cdot \mathbb{G}^{R}(d_R) \cdot\underrightarrow{\mathcal{Y}}^{(m)} ,
\end{eqnarray}
where $d_L$ and $d_R$ are the depths of the left and right gratings, respectively.
The interpretation of \eqref{CasOp2} is particularly simple in terms of modes. Indeed, as we discussed above,  the determinant of the pivotal matrix gives the resonance of the system, and hence 
$\mathrm{det}\left[\Theta^{L}_{22} \cdot \underrightarrow{\mathcal{P}}^{(v)}(-a)\cdot\Theta^{R}_{22}\right] =
\mathrm{det}\left[\Theta^{L}_{22}\right] \mathrm{det}\left[\underrightarrow{\mathcal{P}}^{(v)}(-a)\right]\mathrm{det}\left[\Theta^{R}_{22}\right]$
gives the resonances of the two isolated gratings. 
The factor  $\mathrm{det}\left[\underrightarrow{\mathcal{P}}^{(v)}(-a)\right]=e^{a\sum_{\nu} \lambda_{\nu}}$ represents the contribution of  the continuum of electromagnetic vacuum modes hitting the gratings 
\cite{Intravaia12a}. 
Similarly, the determinant of
$\Theta^{L}_{21}\cdot \underleftarrow{\mathcal{P}}^{(v)}(-a)\cdot \Theta^{R}_{12}+\Theta^{L}_{22}\cdot\underrightarrow{\mathcal{P}}^{(v)}(-a)\cdot\Theta^{R}_{22} $
gives the coupled modes of the two gratings. Indeed, one can show that this matrix is the pivotal matrix $\Theta_{22}^{\rm comp}$ of the composite system formed by the left grating, the vacuum region, and the right grating. This is particularly evident if one writes it as follows
\begin{equation}
\Theta^{\rm comp}_{22} = \underrightarrow{\mathcal{Y}}^{(m)\dag}\cdot \mathbb{G}^{L}(h_{L}) \cdot \mathbb{G}^{(v)}(a) \cdot \mathbb{G}^{R}(h_{R})\cdot\underrightarrow{\mathcal{Y}}^{(m)} ,
\end{equation}
where
\begin{equation}
\mathbb{G}^{(v)}(a)=\underleftarrow{\mathcal{Y}}^{(v)}\cdot \underleftarrow{\mathcal{P}}^{(v)}(-a) \cdot \underleftarrow{\mathcal{Y}}^{(v)\dag}+\underrightarrow{\mathcal{Y}}^{(v)}\cdot \underrightarrow{\mathcal{P}}^{(v)}(-a)\cdot \underrightarrow{\mathcal{Y}}^{(v)\dag}
\label{vacuum-propagator}
\end{equation}
is the  vacuum (propagator) operator. In the limit $a \rightarrow \infty$ (infinitely separated gratings) the second term in the numerator of
\eqref{CasOp2} vanishes, and hence we can write \eqref{pressure} as
\begin{align}
P(a)&= -\frac{4}{\beta} \sum^{\infty'} _{l=0}  
\int_{0}^{\infty} \hspace{-0.3cm} dk_y \int_{0}^{\pi/p}\hspace{-0.3cm} d\alpha_0\,
\partial_{a}\log \frac{\mathrm{det} \left[\Theta^{\rm comp}_{22}(a) \right]}{\mathrm{det}\left[\Theta^{\rm comp}_{22}(a\to \infty)\right]} \nonumber\\
&= \frac{4}{\beta} \sum^{\infty'} _{l=0}  
\int_{0}^{\infty} \hspace{-0.3cm}dk_y \int_{0}^{\pi/p}\hspace{-0.3cm} d\alpha_0\,
\partial_{a}\left.\log \mathrm{det} \left[\underrightarrow{\mathcal{T}}^{\rm comp} \right]\right|^{a}_{a\to \infty} ,
\label{PressureFinal}
\end{align}
where we used the definition of the transmission operator in terms of the pivotal matrix given in eq.\eqref{scattOps}. 

Before concluding this subsection let us discuss the impact
of mode degeneracy on the Casimir interaction. In the case of two degenerate eigenvalues new expressions for the eigenfunctions $\mathcal{U}^{(s)}$ must be found using standard techniques.
Although possible and not mathematically involved, this is however of no use for the evaluation of the Casimir pressure. Indeed, one can show that this will require the modification of the integrand of the previous expression in 
a Lebesgue null measure ensemble of points, without changing the final result.


\subsection{Generalization to non-lamellar gratings}

The simple physical reasoning behind the previous results allows us to generalize the calculation to multilayered periodic structures and non-lamellar gratings, which can be approximated,
by slicing them, as multilayered periodic structures of individual lamellar gratings \cite{LiMultilayer93,MOharam95}. Consider, for example, two non-lamellar gratings
facing each other and separated by vacuum. The composite system is bounded by two homogeneous bulk media $m_L$ and $m_R$.
The pivotal matrix for the composite system is clearly given by
\begin{equation}
\Theta^{\rm comp}_{22}= \underrightarrow{\mathcal{Y}}^{(m_L)\dag}\cdot(\prod_{i} \mathbb{G}^{L}_i )\cdot\mathbb{G}^{(v) }\cdot(\prod_{j}\mathbb{G}^{R}_j)\cdot\underrightarrow{\mathcal{Y}}^{(m_{R})} ,
\end{equation}
where $\mathbb{G}^{L,R}_i$ is the propagator for the $i$-th lamellar slice of the non-lamellar left or right grating. 


\subsection{Large distance asymptotic expression}

Despite the complexity of the previous expressions, it is possible to derive a close expression for the Casimir free energy between gratings in the
asymptotic limit of large distances, $a\rightarrow \infty$.  Since the operator $\underrightarrow{\mathcal{P}}(a)$ is a diagonal matrix with
decreasing exponentials $e^{- a \lambda_{\nu}}$ as matrix elements, it follows that, for any fixed Matsubara frequency, the eigenvalue with $\nu=0$ is the
one that gives the slowest decrease as $a$ grows. This value of $\nu$ corresponds to the zeroth order of reflection in the standard Rayleigh
formalism for scattering from periodic structures. Therefore,  at large distances we can keep only contributions arising from the $\lambda^{(v)}_{\nu=0}$ eigenvalue, and approximate
$1-  \underleftarrow{\mathcal{R}}^{L} \cdot \underrightarrow{\mathcal{P}}(a) \cdot \underrightarrow{\mathcal{R}}^{R} 
\cdot \underrightarrow{\mathcal{P}}(a)
\approx
1- [\underleftarrow{\mathcal{R}}^{L}]_{00} \cdot [\underrightarrow{\mathcal{R}}^{R}]_{00} \; e^{-2\kappa a}$,
where $\kappa=\sqrt{\xi^2 + k_{y}^{2} +\alpha_{0}^{2}}$. The subscript 00 indicates that only the $\gamma=\nu=0$ block of the $\mathcal{R}$ matrices is considered.
For the same reason, at distances large enough the dominant contributions to the Casimir free energy comes from $\alpha_{0}\approx 0 $ and $k_{y} \approx 0$. We know already that when $k_{y}\to 0$ the $e$ and $h$ polarizations decouple, which implies that the submatrices 
$[\underleftarrow{\mathcal{R}}^{L}]_{00}$ and $[\underrightarrow{\mathcal{R}}^{R}]_{00}$ become diagonal. Hence, in this large distance limit, we
approximate the pressure as
\begin{align}
P(a) & \approx  -\frac{4}{\beta} \sum^{\infty'} _{l=0}  
\int_{0}^{\infty} \hspace{-0.3cm}dk_y \int_{0}^{\pi/p} \hspace{-0.3cm}d\alpha_0  \nonumber \\
&\times \partial_{a}\left\{
\log\left[1-[\underleftarrow{\mathcal{R}}^{L}]^{(ee)}_{00}[\underrightarrow{\mathcal{R}}^{R}]^{(ee)}_{00}e^{-2\kappa a}\right] \right. \nonumber\\
&  \left. + \log\left[1-[\underleftarrow{\mathcal{R}}^{L}]_{00}^{(hh)}[\underrightarrow{\mathcal{R}}^{R}]_{00}^{(hh)}e^{-2\kappa a}\right] \right\} ,
\end{align} 
which is formally equivalent to the integrand of the Lifshitz formula for parallel planes. 
At large distances, the zeroth Matsubara frequency ($l=0$) dominates the above summation, which implies that
$P(a)$ is proportional to  $- k_{B}T a^{-3}$, as in the plane-plane case.
The proportionality factor depends on the value of the reflection amplitudes in the limit $k_{y} \approx 0$ and $\alpha_{0} \approx 0$.
For metallic gratings described by the Drude model, we have seen above (see \eqref{zero-limit-1} and \eqref{zero-limit-2}) 
that $[\mathcal{R}]^{(ee)}_{00}(\xi=0, k_y=0) =0$,
while $[\mathcal{R}]^{(hh)}_{00}(\xi=0,k_y=0,\alpha_{0} \to 0)=1$. 
Therefore, as for Drude parallel plates, the prefactor is $\zeta(3)/8\pi$.


\subsection{Numerical results}

\begin{figure}[hhh]
\includegraphics[width=8.5cm]{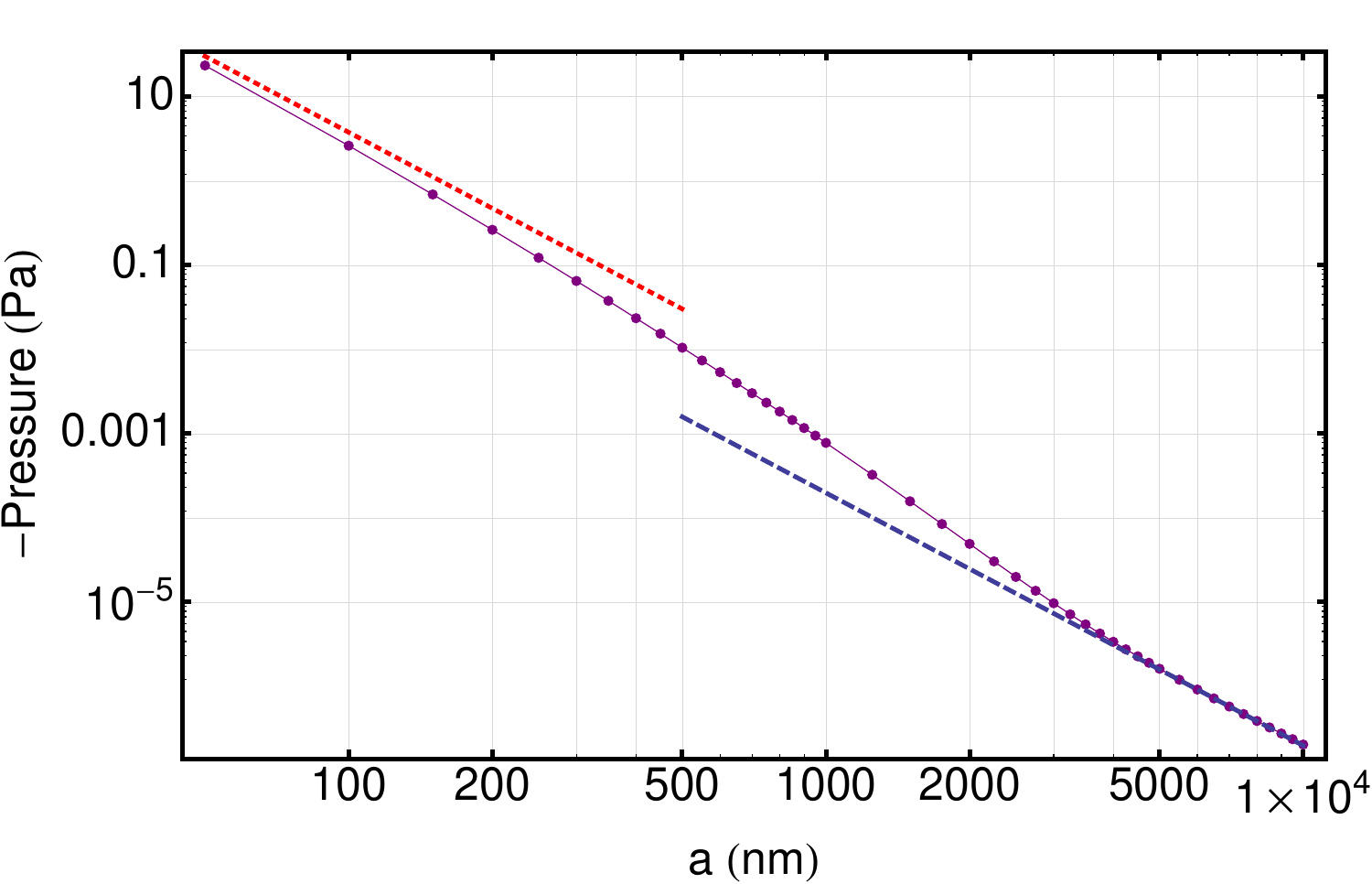}
\caption{Casimir pressure between a metallic grating and a metallic plane, computed using our quasi-analytical modal approach. 
At large separation the pressure tends towards the value $\zeta(3)k_{B}T/(8\pi a^{3})$ (dashed curve). At short separations the pressure is $\propto a^{-3}$ because of the finite grating conductivity. The prefactor used for the dotted curve is the one for the plane-plane case multiplied by the filling factor $f=p_{2}/p$ (see definition of $P_{\rm filling}$ in the text).
The Drude parameters are  $\omega_{p}=8.39$ eV and $\gamma=0.043$ eV. The geometrical parameters of the grating are:  width of the grooves $p_{1}=160$ nm, width of the teeth $p_{2}=90$ nm, and height $d=216$ nm.  Temperature is set to $T=300$ K.
}
\label{Pressure}
\end{figure}

In this subsection we will focus on the Casimir interaction between a gold lamellar grating parallel to a gold flat surface.  
In principle, our modal approach can treat this problem almost fully analytically, requiring numerics only for finding the roots of the transcendental equation
\eqref{transcental-imaginary} to determine the eigenvalues for the grating region. However, from the practical point of view, we are also forced to truncate the matrices and the series, to numerically evaluate integrals, and to deal with convergency issues.  We address these issues in what follows.

The size of the theta- and scattering matrices is set by the number of eigenvectors $N_{\rm max}$ one keeps to describe the fields in the homogeneous regions
(equivalent to the Rayleigh orders). This number will be always odd because we will truncate the Rayleigh expansion symmetrically with respect to the zeroth order. The corresponding matrices will be block matrices with dimension $(2 N_{\rm max}) \times (2 N_{\rm max})$ (the factor $2$ comes from the two polarizations). 
The expression of the grating operator \eqref{grating-propagator}
is formally independent of the truncation order $N_{\rm max}$, and the series defining it could be truncated at a different value, say $M_{\rm max}$. Numerical studies show, however, that for the reflection matrix the best convergency is obtained when  $M_{\rm max}= N_{\rm max}$. This can be physically understood from a argument of dimensionality matching between the Hilbert spaces describing the field inside the grating and in the homogeneous regions. This is particular clear at high frequency where a one-to-one correspondence between grating eigenvectors and vacuum eigenvectors is required to
satisfy the high-frequency transparency.  
Our numerical studies show that, for our choice of optical and geometrical parameters (plasma frequency 8.39 eV, dissipation rate 0.043 eV,  $p_{1}=160$ nm, $p_{2}=90$ nm, $d=216$ nm)  the first eleven modes ($M_{\rm max}= N_{\rm max}=11$) for the $e$- and for $h$-polarization are enough for the theta-matrices (and, hence, the reflection matrices)  to converge for all values of $\xi$, $k_y$, and $\alpha_0$ relevant in the numerics. For our configuration, higher modes would correspond to values much larger than the plasma frequency, for which the metal is almost transparent (see fig.\ref{PlotEigen}). Since the magnitude of the eigenvalues decreases with the inverse of the grating parameters (see \eqref{solh1}, \eqref{solh2}, and \eqref{sole}), more (less) modes will be required for gratings with larger (smaller) geometrical features.

The calculation of the Casimir pressure in eq.\eqref{PressureFinal} also requires the evaluation of two integrals over wave vectors.
The integration is performed using a 30 points Gauss-Legendre quadrature scheme for $\alpha_{0}$,  and a 20 points Gauss-Laguerre quadrature scheme for 
$k_{y}$.  Numerical checks show that for the zeroth Matsubara frequency the agreement with a Montecarlo calculation is better that  1 \% for  
$100 \, {\rm nm} \le a \le 5 \, \mu$m,  and better that 3 \% for $5 \, \mu {\rm m} \le a \le 10 \, \mu$m. The agreement greatly improves for higher Matsubara frequencies. The Matsubara series was truncated at 41 terms. At the distance of $a=50$ nm, the total result changes by less than 
1 \% in going from 37 to 41 Matsubara terms.

Figure \ref{Pressure} shows the result  of the numerical evaluation of the Casimir pressure obtained from \eqref{PressureFinal}. As a check of our prediction we also show the large distance asymptotic expression discussed in the previous section (dashed line). The  dotted line represents the short distance plane-plane asymptotic behavior multiplied by the filling factor ($f=p_{2}/p$) \cite{Intravaia07},
$P_{\rm filling}(a) \equiv - 1.79 f \omega_{p} \hbar c\pi / 720 a^{3}$.
The good agreement between the full line and the dotted line in Fig.  \ref{Pressure} indicates that at short distances the plane-grating Casimir pressure is substantially less than the plane-plane pressure mainly due to geometrical effects.

\begin{figure}
\includegraphics[width=8.5cm]{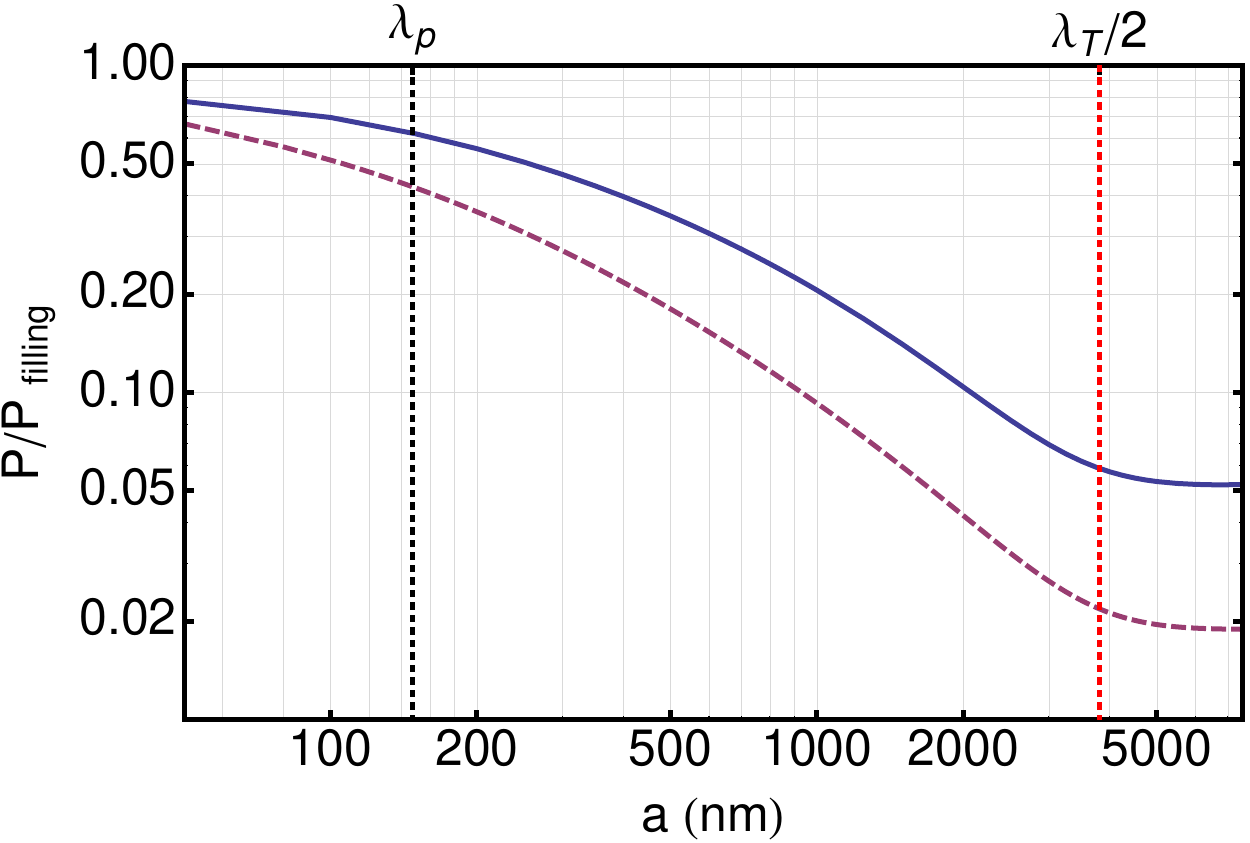}
\caption{
Plane-plane (dashed) and plane-grating (solid) Casimir pressure normalized by $P_{\rm filling}$ with the respective filling factor ($f=1$ for the plane-plane and $f=90/250$ for the plane-grating). The two vertical lines are located at  distances corresponding to the plasma wave-length ($\lambda_{p}=2\pi/\omega_{p}$) and half of the thermal wave-lenght ($\lambda_{T}=\hbar c/ 2k_{B}T $).
The transition from the $a^{-3}$ non-retarded behavior to the $a^{-4}$ retarded behavior for the plane-plane configuration occurs much faster than for the plane-grating one. The large distance $a^{-3}$ thermal regime is however not affected by the grating and starts roughly at the same point for both configurations. Parameters are the same as in previous figures.
 }
\label{pshortplot}
\end{figure} 

Finally, we briefly address the influence of finite conductivity and temperature for Casimir interactions involving gratings. In the plane-plane configuration the Casimir pressure goes as $a^{-3}$ for $a\ll\lambda_{p}$ (non-retarded van der Waals regime), as $a^{-4}$ for $\lambda_{p}\ll a \ll \lambda_{T}$ (retarded regime), and again as $a^{-3}$ for $a \gg \lambda_{T}$ (thermal regime), where $\lambda_{p}=2\pi c/\omega_{p}$ is the plasma wavelength ($\approx 147$ nm in our case) and $\lambda_{T}=\hbar c/ k_{B}T $ is the thermal wavelength ($\approx 7$ $\mu$m at $T=300$ K). 
The behavior at short distances can also be interpreted as resulting from the non-retarded interaction between surface plasmons \cite{Van-Kampen68, Genet04,Intravaia05,Intravaia07,Haakh10}. On the other hand, a grating is known for modifying the electromagnetic near field, by affecting the behavior of surface plasmon modes and effectively increasing the plasma wavelength (as we discussed above).
Therefore, one expects a wider non-retarded regime for the case of metallic gratings.
Figure \ref{pshortplot} shows the grating-plane and the plane-plane pressure normalized by $P_{\rm filling}$
with the respective filling factors ($f=1$ for the plane-plane configuration).
As expected,  the transition from the $a^{-3}$ to the $a^{-4}$ behavior happens at a larger distance for 
the plane-grating than for the plane-plane configuration (i.e., the grating-plane case has a wider non-retarded regime). 
The same figure also shows that, on the contrary, the thermal regime is not affected and starts roughly at the same point for both structures.


\section{Conclusions}

In summary, we have developed a quasi-analytical modal approach to computing Casimir interactions involving 1D lamellar gratings. The method
can be generalized to more complex nanostructures by approximating them via slicing as a collection of multilayered lamellar gratings
 \cite{LiMultilayer93,MOharam95}. 
The key features of our method is that the eigenmodes of the grating can be solved for analytically, while the eigenfrequencies are solutions to a simple transcendental equation \eqref{transcental-imaginary}. Apart from these fundamental aspects, we have also presented an approach to calculate the Casimir
interaction without resorting to any matrix inversion that avoids several potential numerical instabilities, improves the precision of the numerical
results, and can be used in other non-modal frameworks. 
We studied analytically the form of the eigenvalues in some specific limiting cases, and discussed
their impact on the scattering operators and on the Casimir interaction. By analyzing the mode structure at real frequencies, this formalism can also be applied to study other fluctuation-induced interactions (thermal emission, near-field heat transfer, etc.).


\section{Aknowledgments}
We thank R. O. Behunin and R. Gu{\'e}rout for interesting discussions related to this work.
This work was partially supported by the LANL LDRD program and the DARPA/MTOÕs Casimir Effect Enhancement program under DOE/NNSA Contract No. DE-AC52-06NA25396 and DOE-DARPA MIPR 09-Y557. This work was performed, in part, at the Center for Nanoscale Materials, a U.S. Department of Energy, Office of Science, Office of Basic Energy Sciences User Facility under Contract No. DE-AC02-06CH11357. RD thanks the Integrated Nanosystems Developement Institute and the Indiana University Collaborative Research Grants.




\begin{thebibliography}{10}

\bibitem{Davids10}
P.~S. Davids, F. Intravaia, F.~S.~S. Rosa, and D.~A.~R. Dalvit, Phys. Rev. A
  {\bf 82},  062111  (2010).

\bibitem{Casimir48}
H.~B.~G. Casimir, Proc. K. Ned. Ak. Wet. {\bf 51},  793  (1948).

\bibitem{Buscher04}
R. B{\"u}scher and T. Emig, Phys. Rev. A {\bf 69},  062101  (2004).

\bibitem{Lambrecht08}
A. Lambrecht and V.~N. Marachevsky,  Phys. Rev. Lett. {\bf 101}, 160403 (2008).

\bibitem{Johnson11}
S. Johnson,  in {\em Casimir Physics}, Vol. {\bf 834} of {\em Lecture Notes in
Physics}, edited by D.A.R. Dalvit, P.W. Milonni, D.C. Roberts, and F.S.S. Rosa
(Springer, Heidelberg, 2011), pp.\ 175--218.

\bibitem{Rodriguez11}
A.~W. Rodriguez, F. Capasso, and S.~G. Johnson, Nat. Photon. {\bf 5},  211
  (2011).

\bibitem{Ninham70}
B.~W. Ninham, V.~A. Parsegian, and G.~H. Weiss, J. Stat. Phys. {\bf 2},  323
  (1970).

\bibitem{Dalvit06}
D.~A.~R. Dalvit, F.~C. Lombardo, F.~D. Mazzitelli, and R. Onofrio, Phys. Rev. A
  {\bf 74},  020101  (2006).

\bibitem{Intravaia07}
F. Intravaia, C. Henkel, and A. Lambrecht, Phys. Rev. A {\bf 76},  033820
  (2007).

\bibitem{Lambrecht06}
A. Lambrecht, P.~A.~M. Neto, and S. Reynaud, New J. Phys. {\bf 8},  243
  (2006).

\bibitem{Rahi09}
S.~J. Rahi, T. Emig, N. Graham, R.~L. Jaffe, and M. Kardar, Phys. Rev. D {\bf
  80},  085021  (2009).

\bibitem{Rahi11}
S. Rahi, T. Emig, and R. Jaffe,  in {\em Casimir Physics}, Vol. {\bf 834} of {\em
Lecture Notes in Physics}, edited by D.A.R. Dalvit, P.W. Milonni, D.C. Roberts, and
F.S.S. Rosa (Springer, Heidelberg, 2011), pp.\ 129--174.

\bibitem{Lambrecht11a}
A. Lambrecht, A. Canaguier-Durand, R. Gu{\'e}rout, and S. Reynaud,  in {\em
Casimir Physics}, Vol. {\bf 834} of {\em Lecture Notes in Physics}, edited by D.A.R.
Dalvit, P.W. Milonni, D.C. Roberts, and F.S.S.Rosa (Springer, Heidelberg, 2011), pp.\ 97--127.

\bibitem{Rodriguez09}
A.~W. Rodriguez, A.~P. McCauley, J.~D. Joannopoulos, and S.~G. Johnson, Phys.
  Rev. A {\bf 80},  012115  (2009).

\bibitem{Busch07}
K. Busch, G. von Freymann, S. Linden, S. Mingaleev, L. Tkeshelashvili, and M.
  Wegener, Physics Reports {\bf 444},  101   (2007).

\bibitem{Intravaia05}
F. Intravaia and A. Lambrecht, Phys. Rev. Lett. {\bf 94},  110404  (2005).

\bibitem{Haakh10}
H. Haakh, F. Intravaia, and C. Henkel, Phys. Rev. A {\bf 82},  012507  (2010).

\bibitem{Guerout12}
R. Gu{\'e}rout {\it et~al.}, Phys. Rev. B {\bf 85},  180301  (2012).

\bibitem{Cole68}
R. Cole, {\em Theory of ordinary differential equations}
(Appleton-Century-Crofts, New York, 1968).

\bibitem{Naimark68}
M. Naimark,  in {\em Linear differential operators, Part 1}, edited by W.
Everitt (New York: Ungar, New York, 1968).

\bibitem{Botten81b}
I.~C. Botten, M.~S. Craig, R.~C. McPhedran, J.~L. Adams, and J.~R. Andrewartha,
  Opti. Acta {\bf 28},  413  (1981).

\bibitem{Li93}
L. Li, J. Mod. Optics {\bf 40},  553  (1993).

\bibitem{Matsubara55}
T. Matsubara, Prog. Theor. Phys. {\bf 14},  351  (1955).

\bibitem{Carniglia71}
C.~K. Carniglia and L. Mandel, Phys. Rev. D {\bf 3},  280  (1971).

\bibitem{Jackson75}
J. Jackson, {\em Classical Electrodynamics} (John Wiley and Sons Inc., New
  York, 1975).

\bibitem{Intravaia09}
F. Intravaia and C. Henkel, Phys. Rev. Lett. {\bf 103},  130405  (2009).

\bibitem{Suratteau83}
J.~Y. Suratteau, M. Cadilhac, and R. Petit, J. Optics {\bf 14},  273  (1983).

\bibitem{Leeuwen21}
H.-J. van Leeuwen, J. Phys. Radium {\bf 2},  361  (1921).

\bibitem{Bimonte09}
G. Bimonte, Phys. Rev. A {\bf 79},  042107  (2009).

\bibitem{Intravaia12a}
F. Intravaia and R.O. Behunin, in preparation.
  
\bibitem{LiMultilayer93}
L. Li, J. Opt. Soc. Am. A {\bf 10}, 2581 (1993).

\bibitem{MOharam95}
M.G. Moharam {\it et. al.}, J. Opt. Soc. Am. A {\bf 12}, 1077 (1995).

\bibitem{Van-Kampen68}
N. van Kampen, B. Nijboer, and K. Schram, Phys. Lett. A {\bf 26},  307  (1968).

\bibitem{Genet04}
C. Genet, F. Intravaia, A. Lambrecht, and S. Reynaud, Ann. Fond. L. De Broglie {\bf 29},  311  (2004).


\end{thebibliography}
\end{document}